\documentclass[10pt,a4paper,english,prb,aps,twocolumn,superscriptaddress]{revtex4-1}
\usepackage[T1]{fontenc}
\usepackage[latin9]{inputenc}
\usepackage{color}
\usepackage{verbatim}
\usepackage{bm}
\usepackage{amsmath}
\usepackage{amssymb}
\usepackage{graphicx}
\usepackage{wasysym}

\makeatletter

\pdfpageheight\paperheight
\pdfpagewidth\paperwidth

\usepackage{amssymb}
\usepackage{amsmath}
\usepackage{graphicx}
\usepackage[colorlinks=true,linkcolor=blue,anchorcolor=red,citecolor=blue,urlcolor=blue]{hyperref}

\makeatother

\usepackage{babel}
\begin{document}
\renewcommand{\figurename}{Fig.}
\title{Quantum oscillations in acoustic phonons in Weyl semimetals}
\author{Song-Bo Zhang}
\affiliation{Institut f\"ur Theoretische Physik und Astrophysik, Universit\"at
W\"urzburg, D-97074 W\"urzburg, Germany}
\author{Jianhui Zhou}

\affiliation{Anhui Province Key Laboratory of Condensed Matter Physics at Extreme
Conditions, High Magnetic Field Laboratory,Chinese Academy of Sciences,
Hefei 230031, Anhui, China}
\date{\today}
\begin{abstract}
We theoretically study the modification of the energy spectrum of long-wavelength acoustic phonons due to the electron-phonon interaction in a three-dimensional topological Weyl semimetal under the influence of quantizing magnetic fields. We find that the dispersion and attenuation of phonons show striking oscillatory behaviors when varying the magnetic field at low temperatures. These oscillations are distinct when the Fermi energy is in different energy regimes. Moreover, the van Hove singularity of the Weyl spectrum can manifest as a transition between different oscillation patterns when increasing the magnetic field. These phonon behaviors could provide testable fingerprints of the Fermi-surface morphology and relativistic feature of the Weyl semimetal.
\end{abstract}
\maketitle

\section{Introduction}

Three-dimensional (3D) Weyl semimetals are a novel topological phase
of matter in which two low-energy bands have isolated and linear
crossing points, termed as Weyl nodes, in momentum space~\citep{Hosur13Physique,Weng2016jpcm,Burkov2016Nmat,HZLu17review,YanBH2017arcmp,Armitage18RMP}.
According to the fermion doubling theorem~\citep{Nielsen81npb},
these Weyl nodes appear in pairs with opposite chirality. They act as
sources of monopoles and anti-monopoles in momentum space~\citep{Xiao10RMP}.
This nontrivial momentum-space topology results in various fascinating
phenomena, such as the chiral anomaly~\citep{Nielsen83plb,Aji12prbrc,Liu13prb,Goswami13prb},
chiral magnetic effect~\citep{Fukushima08PRD,Zyuzin12prb,Zhou13CPL,Vazifeh13prl,Chang15PRB,Ma2015PRBCME,ZhongSD2016PRLGME},
anomalous magnetoresistance~\citep{Son13prb,Burkov14prl,Gorbar14prb,Lu15Weyl-shortrange}
as well as exotic phonon dynamics \citep{Lundgren15PRB,SongZD2016PRB,LiuDH2017PRL,Rinkel2017PRL,Laliberte19arXiv,Rinkel19PRB}.
The Weyl semimetal breaks either time-reversal or inversion symmetry.
However, under the protection of some extra symmetries, the Dirac
semimetal, which may be simply regarded as a combination of Weyl semimetals
related by time-reversal or inversion symmetry, could exist and inherit
many interesting properties of Weyl semimetals. Very recently, Weyl and Dirac
semimetals have been predicted in a variety of materials~\citep{TangF19Nature,Vergniory19Nature,ZhangTT19Nature},
some of which have been confirmed experimentally~\citep{Hasan2017arcmp,Armitage18RMP}.

Applying a magnetic field in a 3D metal quantizes the continuous electronic
states to a series of discrete Landau bands. The corresponding electronic
orbit (also called cyclotron orbit) at a given energy $\epsilon$ is
closed. In momentum space, it encloses an area $S(\epsilon)$ normal
to the magnetic field. This area obeys the condition of Lifshitz-Onsager
quantization $S(\epsilon)=(2\pi eB/\hbar)(\nu+\gamma_{\text{L}})$~\citep{Shoenberg1984},
where $B$ is the magnetic field and $\nu$ is a non-negative integer.
The phase mismatch $0\leqslant\gamma_{\text{L}}<1$ consists of two
parts: $\gamma_{\text{L}}=\gamma_{\text{M}}+\gamma_{\text{B}}$. The
first part $\gamma_{M}=1/2$ is the Maslov index and the other part
is directly related to the Berry phase $\phi_{\text{B}}$ of the closed
orbit by $\gamma_{\text{B}}=-\phi_{\text{B}}/2\pi$~\citep{Fuchs10EPL}.
As the result of the quantization, many measurable quantities oscillate
when varying the magnetic field at low temperatures, such as the magnetization
and longitudinal conductivity associated with the de Haas-van Alphen
and Shubnikov-de Haas effect, respectively. Several experiments on
the Dirac semimetals Cd$_{3}$As$_{2}$\ \citep{He14prl,Zhao15prx,Xiang15PRL,Narayanan15prl,LiH16natcom},
Na$_{3}$Bi\ \citep{Xiong15sci,Xiong16EPL} and the TaAs-class Weyl
semimetals\ \citep{HuangXC15prx,Luo15PRB,ZhangCL16natcom,WangZ15arXiv}
have made efforts to detect the relativistic feature of Weyl fermions.
Very recently, people have commenced to experimentally explore the
behaviors of phonons in TaAs in the presence of magnetic fields \citep{Xiang2019PRX}.
Therefore, theoretical investigations on the phonon physics in Weyl
semimetals under magnetic fields are highly desired.

In this paper, we study the modification of the energy spectrum
of long-wavelength acoustic phonons by the electron-phonon interaction
in doped Weyl semimetals under quantizing magnetic fields at low temperatures.
We find that when the Fermi energy lies in different energy regimes,
the modification of phonon dispersion $\Delta\Omega_{q}$ shows distinct
oscillations when varying the magnetic field. When the Fermi energy
is close to the Weyl nodes, the oscillations correspond to a single
period which resembles that of the ideal Weyl fermions. However, due
to the higher-order momentum terms, the phase shift in the oscillations
differs from that of the ideal Weyl fermions by a value proportional
to the Fermi energy measured from the Weyl nodes. When the Fermi
energy lies above the Lifshitz transition energy, multiple periods
could coexist in $\Delta\Omega_{q}$, resulting in beating oscillations.
Moreover, a van Hove singularity could exist in the Weyl spectrum
and be signified as a transition between different patterns of oscillations.
{} In addition, we show that the phonon attenuation exhibits sets of
periodic spike peaks at low temperatures the features of which are also very
sensitive to the position of the Fermi energy. These
findings could greatly enrich our understanding of Weyl semimetals
as well as their coupling to acoustic phonons.

The paper is organized as follows. In Sec.\ \ref{sec:Model-and-formalism},
we introduce the minimal model for topological Weyl semimetals and
outline the approach to calculate the modification of the energy dispersion
and the attenuation of long-wavelength acoustic phonons in the Weyl
semimetal under magnetic fields. In Sec.\ \ref{sec:Renormalization-of-phonon},
we analyze the modification of phonon dispersion at zero temperature
in two complementary parameter regimes, respectively. In Sec.\ \ref{sec:Phonon-attenuation-for},
we calculate and discuss the phonon attenuation at low temperatures.
Section\ \ref{sec:Summary} summarizes the results.

\section{Model and formalism\label{sec:Model-and-formalism}}

\begin{figure}[htp]
\centering \includegraphics[width=8.5cm]{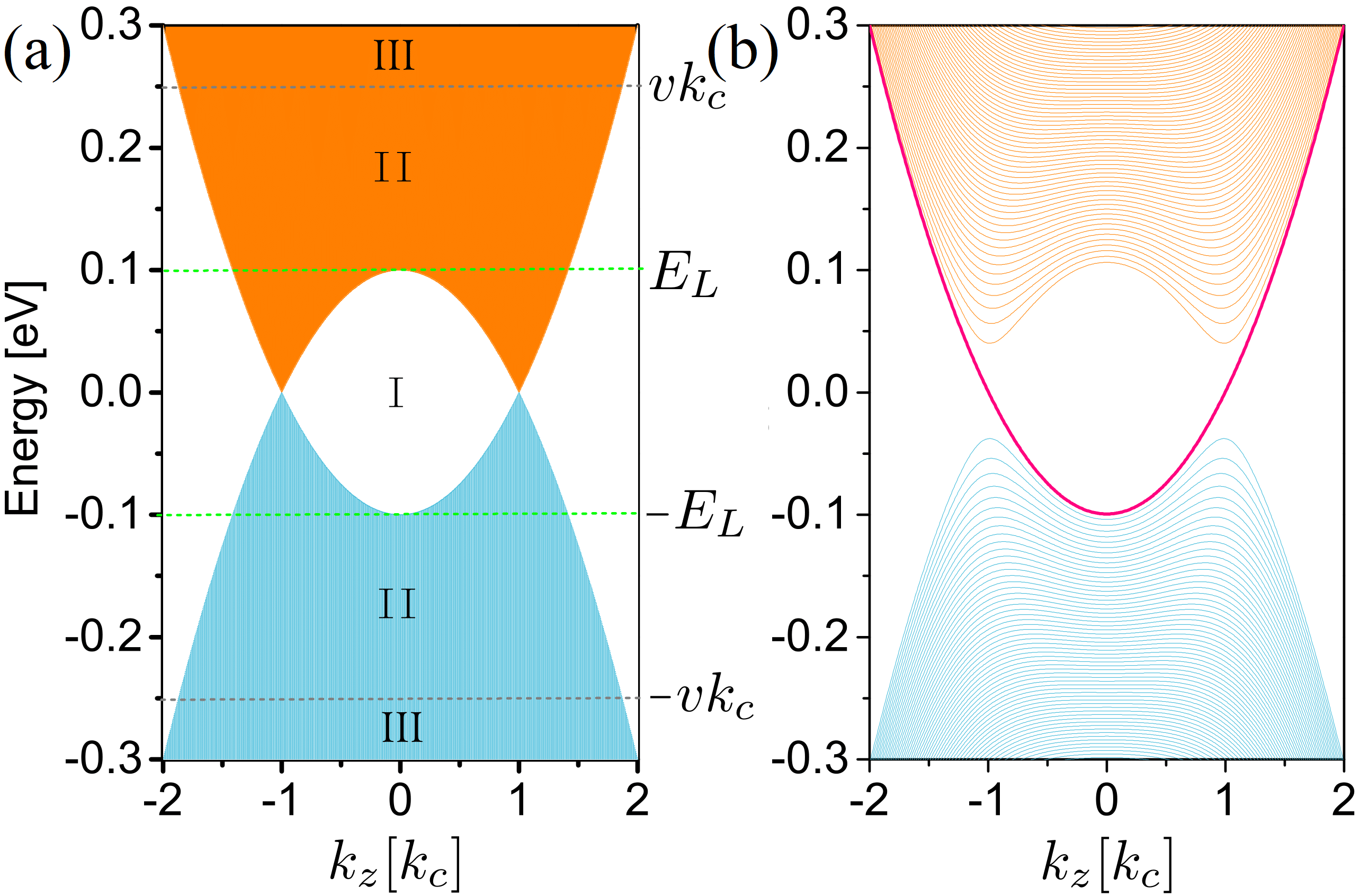} \caption{(a) Energy spectrum of the model \eqref{eq:model_hamiltonian} with respect to $k_{z}$ in the absence of the magnetic field. The dotted lines
mark the boundaries between three energy regimes. (b) Landau bands
with respect to $k_{z}$ in the presence of an $8$ T magnetic field
in the $z$ direction. The pink thick curve stands for the chiral
Landau band. Other parameters are $k_{c}=0.1$ $\mathring{\text{A}}$$^{-1}$, $m=10$
eV$\cdot$$\mathring{\text{A}}$$^{2}$ and $v=2.5$ eV$\cdot$$\mathring{\text{A}}$.}

\label{Fig:1LL}
\end{figure}

The minimal model for topological Weyl semimetals is given by~\citep{Lu15Weyl-shortrange}
\begin{eqnarray}
\mathcal{H}({\bf k}) & = & v(k_{x}\sigma_{x}+k_{y}\sigma_{y})+m(k_{c}^{2}-|{\bf k}|^{2})\sigma_{z},\label{eq:model_hamiltonian}
\end{eqnarray}
where $\sigma_{x,y,z}$ are Pauli matrices acting on the two involved
orbits; $v\equiv v_{F}\hbar$ is the Fermi velocity in the $x$-$y$
plane and ${\bf k}=(k_{x},k_{y},k_{z})$ is the 3D momentum. The model
has two energy bands touching at two Weyl nodes at $(0,0,\pm k_{c})$,
respectively. Around the Weyl nodes, the model mimics two Weyl Hamiltonians
with opposite velocities $\pm v_{z}=\pm2mk_{c}$ in the $z$ direction.
The sign $\pm$ indicates opposite chirality of the two Weyl nodes.
The momentum-quadratic term $m|{\bf k}|^{2}$ is crucial to derive
the Weyl nodes as well as well localized surface states in the presence
of a surface. It breaks the Lorentz symmetry and modifies the linear
dispersion of the Weyl fermions away from the Weyl nodes. The model
 has a Lifshitz transition energy $E_{\text{L}}\equiv mk_{c}^{2}$
at which the two separated Fermi surfaces merge to a single one; see Fig.\ \ref{Fig:1LL}(a). When
$v_{z}>\sqrt{2}v$, the model possesses also a van Hove singularity
at $E_{\text{v}}\equiv v\sqrt{v_{z}^{2}-v^{2}}/m$ where the density
of states (DOS) diverges. The model along with its time-reversal partner
has been used to well describe the low-energy physics of topological
Dirac semimetals such as Cd$_{3}$As$_{2}$ and Na$_{3}$Bi\ \citep{Wang12prb,Wang13prb}.

In this paper, we consider a uniform magnetic field $B$ applied in the $z$ direction to the Weyl semimetal \cite{footNote2}. The electronic spectrum of the Weyl semimetal is split to a series
of Landau bands~\citep{Lu15Weyl-shortrange,ZhangSB16NJP}
\begin{eqnarray}
\epsilon_{k_{z}}^{0} & = & \omega-\mathcal{M}_{0}(k_{z}),\nonumber \\
\epsilon_{k_{z}}^{ns} & = & \omega+s\sqrt{\mathcal{M}_{n}^{2}(k_{z})+n\eta^{2}},~~n\geqslant1\label{eq:Weyl_LLs}
\end{eqnarray}
where $\eta=\sqrt{2}v/\ell_{B}$, $\omega=m/\ell_{B}^{2}$, $\mathcal{M}_{n}(k_{z})=E_{\text{L}}-mk_{z}^{2}-2n\omega$
with $n\in\{0,1,2,...\}$ and $s=\pm$; see Fig.\ \ref{Fig:1LL}(b). $\ell_{B}=\sqrt{\hbar/eB}$
is the magnetic length. All the Landau bands have the degeneracy of
$N_{\text{L}}=L_{x}L_{y}/2\pi\ell_{B}^{2}$ with $L_{x,y,z}$ the
lengths of the system in the $x$, $y$ and $z$ direction, respectively.
The $\omega$ term in the Landau bands stems from the quadratic term
in the model. It breaks the particle-hole symmetry between the $n\geqslant1$ Landau bands. This term plays an important role in the transition
of the patterns of quantum oscillations, as we show below. To facilitate
the derivation, we adopt the Landau gauge $\bm{A}=(-By,0,0)$. Hence,
the wavefunctions of the $n\geqslant1$ Landau bands can be written
as
\begin{eqnarray}
\langle\bm{r}|n,s,k_{x},k_{z}\rangle & = & \left(\begin{array}{c}
C_{k_{z}\uparrow}^{ns}\phi_{n-1k_{x}k_{z}}(\bm{r})\\
C_{k_{z}\downarrow}^{ns}\phi_{nk_{x}k_{z}}(\bm{r})
\end{array}\right),\label{eq:wavefunction}
\end{eqnarray}
where $C_{k_{z}\uparrow}^{n+}=-C_{k_{z}\downarrow}^{n-}=\cos\theta_{k_{z}}^{n}$,
$C_{k_{z}\downarrow}^{n+}=C_{k_{z}\uparrow}^{n-}=\sin\theta_{k_{z}}^{n}$,
$2\theta_{k_{z}}^{n}=\arctan(\sqrt{n}\eta/\mathcal{M}_{n})$ and
\begin{eqnarray}
\phi_{nk_{x}k_{z}}(\bm{r}) & = & \dfrac{e^{ik_{x}x+ik_{z}z}e^{-(y-y_{0})^{2}/2\ell_{B}^{2}}}{\sqrt{2^{n}n!\sqrt{\pi}\ell_{B}L_{x}L_{z}}}\mathcal{H}_{n}\left(\frac{y-y_{0}}{\ell_{B}}\right).\label{eq:eigenstate_harmonic}
\end{eqnarray}
$\mathcal{H}_{n}(x)$ is the $n$-th Hermite polynomial and $y_{0}=k_{x}\ell_{B}^{2}$
is the guiding center. Specially, the wavefunction of the chiral Landau
band with $n=0$ reads $\langle\bm{r}|0,k_{x},k_{z}\rangle=(0,\phi_{0k_{x}k_{z}}(\bm{r}))^{T}$.
Note that we focus on the orbital effect and neglect the Zeeman splitting
effect of the magnetic field throughout this paper.

In the presence of the magnetic field, the interaction between the
electrons and acoustic phonons can be written as 
\begin{eqnarray}
H_{\text{int}} & = & \sum_{\alpha,\alpha'}\sum_{\bm{q}}\Gamma_{\alpha\alpha'}^{(0)}(\bm{q})a_{\alpha}^{\dagger}a_{\alpha'}(b_{\bm{-q}}^{\dagger}+b_{\bm{q}}),\label{eq:Interaction-hamiltonian}
\end{eqnarray}
where the symbol $\alpha$ and $\alpha'$ are abbreviations for the
set of quantum numbers $n,$ $s,$ $k_{x}$ and $k_{z}$; $a_{\alpha}^{\dagger}$
($a_{\alpha}$) is the creation (annihilation) operator for an electron
in the quantum state $|\alpha\rangle$ with energy $\epsilon_{\alpha}$;
$b_{\bm{q}}^{\dagger}$ ($b_{\bm{q}}$) is the creation (annihilation)
operator for an acoustic phonon with the momentum $\bm{q}=(q_{x},q_{y},q_{z})$
and energy $\Omega_{q}=v_{p}q$; $v_{p}/\hbar$ is the phonon
velocity; $\Gamma_{\alpha\alpha'}^{(0)}(\bm{q})=\langle\alpha|U(\bm{q})e^{-i\bm{q}\cdot\bm{r}}|\alpha'\rangle$
are the matrix elements of electron-phonon interaction, $U(\bm{q})$
is the electron-phonon perturbation potential. For simplicity, we
consider $U(\bm{q})$ to be independent of the electron state. For
longitudinal acoustic phonons $|U(\bm{q})|^{2}=\hbar^{2}D^{2}q/\text{(}2v_{p}\rho)$,
where $D$ is the deformation potential constant and $\rho$ is the
mass density of the crystal \citep{Ziman01Book2}.
We utilize the simplified scalar form of the electron-phonon interaction and neglect the gauge-field interaction in Eq.\,(\ref{eq:Interaction-hamiltonian}). It should be noted that the full electron-phonon interaction in Weyl semimetals is a $2\times2$ matrix in the pseudospin space, the non-diagonal elements of which could generate electromagnetic gauge fields~\citep{Rinkel2017PRL,Cortijo15PRL,Arjona18PRB}. The gauge field effect is interesting but goes beyond the scope of this paper.

The Dyson's equation for phonon Green's function is given by~\citep{Mahan1990}
\begin{eqnarray}
D^{-1}(\bm{q},\Omega) & = & D_{0}^{-1}(\bm{q},\Omega)-\Pi(\bm{q},\Omega),\label{eq:phonon_GF}
\end{eqnarray}
where $D_{0}(\bm{q},\Omega)=2\Omega_{\bm{q}}/(\Omega^{2}-\Omega_{\bm{q}}^{2})$
is the phonon Green's function in the absence of interaction. $\Pi(\bm{q},\Omega)$
is the phonon self-energy. It can be calculated from a bubble diagram:
\begin{eqnarray}
\Pi(\bm{q},i\omega_{m}) & = & \dfrac{T}{V}\sum_{i\epsilon_{n}}\sum_{\alpha,\alpha'}G(\alpha,i\epsilon_{n})\Gamma_{\alpha\alpha'}(\bm{q})\nonumber \\
 &  & ~~\times G(\alpha',i\epsilon_{n}-i\omega_{m})\Gamma_{\alpha\alpha'}^{(0)*}(\bm{q}),\label{eq:phonon_SE}
\end{eqnarray}
where $G(\alpha,i\epsilon_{n})$ is the Green's function of an electron
in the state $|\alpha\rangle$. $\epsilon_{n}=(2n+1)\pi T$ with integers
$n$ are the Matsubara frequencies for electrons. $T$ is temperature
(the Boltzmann constant $k_{B}=1$ is taken) and $V=L_{x}L_{y}L_{z}$
is the volume of the system. $\Gamma_{\alpha\alpha'}(\bm{q})$ represents
the electron-phonon interaction vertex. In the Weyl semimetal,
 Migdal's theorem is applicable with an accuracy up to the order
of $v_{p}/v$\ \citep{Roy14PRB}. Note that the phonon velocity is
typically a few orders of magnitude smaller than the electron velocity.
One can thus replace $\Gamma_{\alpha\alpha'}(\bm{q})$ and $G(\alpha,i\epsilon_{n})$
with $\Gamma_{\alpha\alpha'}^{(0)}(\bm{q})$ and $G^{(0)}(\alpha,i\epsilon_{n})=(i\epsilon_{n}-\epsilon_{\alpha}+\mu)^{-1}$,
respectively. Summing over the electron frequencies and then carrying
out the analytic continuation $i\omega_{n}\rightarrow\Omega_{q}+i\delta$,
we can write the retarded phonon self-energy as
\begin{eqnarray}
\Pi^{R}(\bm{q},\Omega_{q}) & = & \dfrac{4\pi^{2}\chi}{V}\sum\limits _{\alpha,\alpha'}|\mathcal{F}_{\alpha\alpha'}(\bm{q})|^{2}\dfrac{f(\epsilon_{\alpha})-f(\epsilon_{\alpha'})}{\epsilon_{\alpha}-\epsilon_{\alpha'}-\Omega_{q}-i\delta},\label{eq:phonon_SE2}
\end{eqnarray}
where $\chi=\hbar^{2}qD^{2}/(8\pi^{2}\rho v_{p})$, $f(\epsilon_{\alpha})=[1+\exp((\epsilon_{\alpha}-\mu)/T)]^{-1}$
is the Fermi function, $\delta$ is a positive infinitesimal and $\mathcal{F}_{\alpha\alpha'}(\bm{q})\equiv\langle\alpha|e^{-i\bm{q}\cdot\bm{r}}|\alpha'\rangle$
is the form factor. Using the wavefunctions stated in Eq.\,(\ref{eq:wavefunction}),
the form factor is derived as
\begin{eqnarray}
\mathcal{F}_{\alpha\alpha'}(\bm{q}) & = & D(\bm{q})[C_{k_{z}\uparrow}^{ns}C_{k_{z}'\uparrow}^{n's'}\mathcal{J}_{n-1}^{n'-1}(u)+C_{k_{z}\downarrow}^{ns}C_{k_{z}'\downarrow}^{n's'}\mathcal{J}_{n}^{n'}(u)]\nonumber \\
 &  & \times\delta_{k_{x}',k_{x}+q_{x}}\delta_{k_{z}',k_{z}+q_{z}}, \nonumber \\
\mathcal{J}_{n}^{n'}(u) & = & \rho^{|n-n'|}L_{\text{min(n,}n')}^{|n-n'|}(u)\sqrt{\dfrac{\text{min}(n,n')!}{\text{max}(n,n')!}},
\end{eqnarray}
where $D(\bm{q})=e^{i\varphi_{\bm{q}}(n'-n)}e^{-iq_{y}(y_{0}'+y_{0})/2}e^{-u/2}$,
$u=\ell_{B}^{2}(q_{x}^{2}+q_{y}^{2})/2$, $\varphi_{\bm{q}}=\arctan(q_{y}/q_{x})$
and $L_{N}^{m}(x)$ are the associated Laguerre polynomials. In particular,
for $n=0$, $C_{k_{z}\uparrow}^{0}=0$ and $C_{k_{z}\downarrow}^{0}=1$.

As mentioned before, we are interested in the case where the wavelength
of the acoustic phonon is sufficiently large so that $q\ell_{B}\ll1$
and $q\ll k_{c}$. In this case, $\mathcal{J}_{n}^{n'}(\rho)\approx\delta_{n,n'}$
and $C_{k_{z}+q}^{ns}\approx C_{k_{z}}^{ns}$ can be taken. These,
along with the energy denominator $\epsilon_{\alpha}-\epsilon_{\alpha'}-\Omega_{\bm{q}}$
in Eq.\,(\ref{eq:phonon_SE2}), make only the terms with $s=s'$
important. Under these circumstances, the interband transitions are
forbidden, and the form factor simplifies to
\begin{eqnarray}
|\mathcal{F}_{\alpha\alpha'}(\bm{q})| & = & \delta_{k_{x}',k_{x}}\delta_{k_{z}',k_{z}+q_{z}}\delta_{n,n'}\delta_{s,s'}.\label{eq:form_factor}
\end{eqnarray}
With Eq.\,(\ref{eq:form_factor}) in Eq.\,(\ref{eq:phonon_SE2})
and considering small phonon self-energy for reality, the modification
of phonon dispersion $\Delta\Omega_{q}=\text{Re}[\Pi^{R}(\bm{q},\Omega_{q})]$
and the attenuation of phonons $\Gamma_{q}=-\text{Im}[\Pi^{R}(\bm{q},\Omega_{q})]$
can be written as
\begin{eqnarray}
\Delta\Omega_{q} & = & \dfrac{\chi}{\ell_{B}^{2}}\sum_{n,s}\int dk_{z}\dfrac{f(\epsilon_{k_{z}}^{ns})-f(\epsilon_{k_{z}+q_{z}}^{ns})}{\epsilon_{k_{z}}^{ns}-\epsilon_{k_{z}+q_{z}}^{ns}-\Omega_{q}},\label{eq:Pi_real}\\
\Gamma_{q} & = & \dfrac{\pi\chi}{\ell_{B}^{2}}\sum_{n,s}\int dk_{z}[f(\epsilon_{k_{z}+q_{z}}^{ns})-f(\epsilon_{k_{z}}^{ns})]\nonumber \\
 & \  & \ \ \ \ \ \ \ \ \ \ \ \ \ \ \ \times\delta(\epsilon_{k_{z}}^{ns}-\epsilon_{k_{z}+q_{z}}^{ns}-\Omega_{q}),\label{eq:Pi_imag}
\end{eqnarray}
respectively, where the sum runs over all Landau bands indexed by
$\{n,s\}$. We have considered a large Weyl system $L_{x,y,z}\rightarrow\infty$
and transformed the summation over $k_{z}$ to an integral. The factor $\ell_B^{-2}$ accounts for the Landau degeneracy which stems from the summation over $k_x$. According
to Eqs.\,(\ref{eq:Pi_real}) and (\ref{eq:Pi_imag}), $\Delta\Omega_{q}$ and $\Gamma_{q}$ are non-zero only when
$q_{z}$ is finite, due to the conservation of momentum and energy.
For a given phonon momentum $\bm{q}$, both $\Delta\Omega_{q}$ and
$\Gamma_{q}$ are most pronounced when $\bm{q}\parallel\bm{B}$ whereas
suppressed when $\bm{q}\perp\bm{B}.$ At zero temperature $T=0$, we can
further approximate
\begin{equation}
f(\epsilon_{k_{z}}^{ns})-f(\epsilon_{k_{z}+q_{z}}^{ns})\approx-q\cos\theta v_{k_{z}}^{ns}\delta(\epsilon_{k_{z}}^{ns}-\mu)
\end{equation}
by considering $q\ll k_{z}$, where $v_{k_{z}}^{ns}=\partial\epsilon_{k_{z}}^{ns}/\partial k_{z}$
and $\theta$ is the angle between the directions of the magnetic
field and the phonon propagation. Then, it follows from Eqs.\,(\ref{eq:Pi_real})
and (\ref{eq:Pi_imag}) that
\begin{eqnarray}
\Delta\Omega_{q} & = & -\dfrac{\chi}{\ell_{B}^{2}}\sum_{n,s}\int dk_{z}\delta(\epsilon_{k_{z}}^{ns}-\mu)\dfrac{v_{k_{z}}^{ns}\cos\theta}{v_{k_{z}}^{ns}\cos\theta+v_{p}},\label{eq:Pi_real-1}\\
\Gamma_{q} & = & \dfrac{\pi\chi v_{p}}{\ell_{B}^{2}}\sum_{n,s}\int dk_{z}\delta(\epsilon_{k_{z}}^{ns}-\mu)\delta(v_{k_{z}}^{ns}\cos\theta+v_{p}).\label{eq:Pi_attentuation}
\end{eqnarray}
Note that the phonon attenuation can also be obtained
from $\Delta\Omega_{q}$ by applying the Kramers-Kr$\ddot{\text{o}}$nig
relation. Equations\,(\ref{eq:Pi_real-1}) and (\ref{eq:Pi_attentuation})
indicate that only the electrons around the Fermi surface participate
in the renormalization of phonon dispersion $\Delta\Omega_{q}$ as well as the phonon
attenuation $\Gamma_{q}$. Both $\Delta\Omega_{q}$ and $\Gamma_{q}$ are proportional to the momentum of the long-wave phonon, similar to the undressed phonon spectrum $\Omega_{q}=v_{p}q$. Thus, these modifications can be directly related to the modification of phonon velocity.  Moreover, they show singularities periodically when the condition $v_{k_{z}}^{ns}\cos\theta+v_{p}=0$
is met \cite{Note1} In the following, we focus on the case of parallel propagation with
$\theta=0$, unless specified otherwise. We emphasize
that the main results remain valid when $\theta$ deviates slightly
from zero.

\section{Phonon dispersion modification at zero temperature\label{sec:Renormalization-of-phonon}}

In this section, we show that at zero temperature, $\Delta\Omega_{q}$
exhibits singularity and oscillatory behaviors when varying the magnetic
field. As indicated in Eq.~(\ref{eq:Pi_attentuation}), the singularities
occur when the component of the electron velocity in the field direction
coincides with that of the phonons. Away from the singularities, the phonon
velocity is much smaller than the electron velocity. Thus, $\Delta\Omega_{q}$
can be rewritten as
\begin{eqnarray}
\Delta\Omega_{q} & = & -\dfrac{\chi}{\ell_{B}^{2}}\int dk_{z}\mathcal{D}(k_{z}),\label{eq:Pi_real_limit}
\end{eqnarray}
where
\begin{eqnarray}
\mathcal{D}(k_{z}) & = & \delta(\epsilon_{k_{z}}^{0}-\mu)+\sum_{n=1}^{\infty}\sum_{s=\pm}\delta(\epsilon_{k_{z}}^{ns}-\mu).
\end{eqnarray}
This indicates that $\Delta\Omega_{q}$ is directly related to the
electron DOS at the Fermi energy. It could therefore provide an alternative
way to probe the information encoded in the electron DOS.

Next, we calculate the DOS function $\mathcal{D}(k_{z})$ explicitly
with the Landau bands stated in Eq.\,(\ref{eq:Weyl_LLs}). The oscillation
behavior occurs in the semiclassical limit, i.e., $|\mu|\gg\eta,\omega$,
where many Landau bands cross the Fermi energy. In this limit, $\mathcal{D}(k_{z})$
can be evaluated by using Poisson's summation rule:
\begin{eqnarray}
\mathcal{D}(k_{z}) & = & 2|\widetilde{\mu}|\text{Re}\Big[\sum_{j=0}^{\infty}\int_{0}^{\infty}dn\delta(\varepsilon_{nk_{z}}^{2}-\widetilde{\mu}^{2})\exp(i2\pi jn)\Big]\nonumber \\
 &  & \quad+\dfrac{1}{2}\left[\delta(\varepsilon_{0k_{z}}-\widetilde{\mu})-\delta(\varepsilon_{0k_{z}}+\widetilde{\mu})\right],
\end{eqnarray}
where $\varepsilon_{0k_{z}}=m(k_{z}^{2}-k_{c}^{2})$, $\varepsilon_{nk_{z}}=\sqrt{\mathcal{M}_{n}^{2}(k_{z})+n\eta^{2}}$
with $n\in\{1,2,\cdots\}$, and for convenience we have defined $\widetilde{\mu}\equiv\mu-\omega$.
Eventually, $\mathcal{D}(k_{z})$ can be rewritten as
\begin{eqnarray}
\mathcal{D}(k_{z}) & = & \dfrac{\ell_{B}^{2}|\mu_{-}|}{\mathcal{F}_{k_{z}}}\sum_{l=\pm}\Theta(n_{l,k_{z}})\Big[1+2\sum_{j=1}^{\infty}\cos\left(2\pi jn_{l,k_{z}}\right)\Big]\nonumber \\
 &  & \qquad+\dfrac{1}{2}\left[\delta(\varepsilon_{0k_{z}}-\widetilde{\mu})-\delta(\varepsilon_{0k_{z}}+\widetilde{\mu})\right],\label{eq:Dfunction}
\end{eqnarray}
where $\Theta(x)$ is the Heaviside step function and the index functions
read
\begin{eqnarray}
n_{\pm,k_{z}} & = & \ell_{B}^{2}\dfrac{2m^{2}(k_{c}^{2}-k_{z}^{2})-v^{2}\pm\mathcal{F}_{k_{z}}}{4m^{2}},\label{eq:n-function}\\
\mathcal{F}_{k_{z}} & = & \sqrt{v^{4}-4v^{2}m^{2}(k_{c}^{2}-k_{z}^{2})+4m^{2}\widetilde{\mu}^{2}}.
\end{eqnarray}

We plug Eq.\,(\ref{eq:Dfunction}) into Eq.\,(\ref{eq:Pi_real_limit})
and perform the integration over $k_{z}$. Due to the large value
of $\ell_{B}$, the factor $\cos(2\pi jn_{l,k_{z}})$ is a rapidly
oscillating function of $k_{z}$. The other factors in Eq.\ (\ref{eq:Dfunction})
are smooth in $k_{z}$. We can employ the saddle point approximation
to the integration by expanding $n_{l,k_{z}}$ around its extremal
points $k_{i}$ over which the integration is important \citep{Abrikosov-book}.
According to Eq.\ (\ref{eq:n-function}), $n_{+,k_{z}}$ has three
possible extremal points:
\begin{eqnarray}
k_{i}^{+} & = & 0,\:\pm\sqrt{k_{c}^{2}-\widetilde{\mu}^{2}/v^{2}},
\end{eqnarray}
while $n_{-,k_{z}}$ has only one at $k_{i}^{-}=0$. Thus, the integration
over $k_{z}$ can be recast as
\begin{align}
 & \int dk_{z}\dfrac{\Theta(n_{l,k_{z}})}{\mathcal{F}_{k_{z}}}\cos(2\pi jn_{l,k_{z}})\nonumber \\
= & \sum_{k_{i}^{l}}\dfrac{\Theta(n_{l,k_{i}^{l}})}{\mathcal{F}_{k_{i}^{l}}\sqrt{jn_{l,k_{i}^{l}}''}}\Big|\cos\Big(2\pi jn_{l,k_{i}^{l}}\pm\dfrac{\pi}{4}\Big),
\end{align}
where $n_{l,k_{z}}''\equiv\partial^{2}n_{k_{z}}^{l}/\partial k_{z}^{2}$
and the $+$ $(-)$ sign is taken if $n_{l,k_{i}^{l}}''>0$ $(<0)$.
Finally, $\Delta\Omega_{q}$ can be written as three parts:
\begin{eqnarray}
\Delta\Omega_{q} & =\Delta\Omega_{q}^{\text{osc}}+ & \Delta\Omega_{q}^{\text{b}}+\Delta\Omega_{q}',\label{eq:Pi_real_final}
\end{eqnarray}
where
\begin{align}
\Delta\Omega_{q}^{\text{osc}} & =-2\chi|\widetilde{\mu}|\sum_{l=\pm}\sum_{k_{i}^{l}}\sum_{j=1}^{\infty}\nonumber \\
 & \times\dfrac{\Theta(n_{l,k_{i}^{l}})}{\mathcal{F}_{k_{i}^{l}}\sqrt{jn_{l,k_{i}^{l}}''}}\cos\Big(2\pi jn_{l,k_{i}^{l}}\pm\dfrac{\pi}{4}\Big).\label{eq:oscillation}\\
\Delta\Omega_{q}^{\text{b}} & =-\chi|\widetilde{\mu}|\sum_{l=\pm}\int dk_{z}\dfrac{\Theta(n_{k_{z}}^{l})}{\mathcal{F}_{k_{z}}},\label{eq:Background}\\
\Delta\Omega_{q}' & =\dfrac{\chi}{2\ell_{B}^{2}}\Big[\dfrac{\Theta(E_{L}-\widetilde{\mu})}{\sqrt{m(E_{L}-\widetilde{\mu})}}-\dfrac{\Theta(E_{L}+\widetilde{\mu})}{\sqrt{m(E_{L}+\widetilde{\mu})}}\Big].
\end{align}
The first part $\Delta\Omega_{q}^{\text{osc}}$, consisting of several
sets of harmonic terms, describes the quantum oscillations of the
phonon spectrum. Since the Landau index $n_{k_{z}}$ is related to
the cross section of the Fermi surface at $k_{z}$ by $n_{k_{z}}=\ell_{B}^{2}S(k_{z},\mu)/2\pi$,
these oscillations are essentially governed by the extremal cross sections
of the Fermi surfaces normal to the magnetic field. Therefore, the
oscillations of $\Omega_{q}$ are of the same type of the de Haas-van
Alphen oscillations and can be used to detect the Fermi surface morphology
of electrons. In reality, a finite broadening of Landau bands due to, e.g., impurity
scattering, electron-phonon interaction or thermal excitation at finite temperatures, smears out
higher harmonics and make only the first harmonic important~\citep{Shoenberg1984}.
Therefore, the quantum oscillations are customarily described by only
the first ($j=1$) harmonic term in experiments. In the following,
we use the first harmonic term to represent a set of harmonics for
convenience unless specified otherwise. The second and last parts
$\Delta\Omega_{q}^{\text{b}}$ and $\Delta\Omega_{q}'$ give the background
of the oscillations. In general, $\Delta\Omega_{q}^{\text{b}}$ also
depends on the magnetic field via the modified Fermi energy $\widetilde{\mu}=\mu-\omega$.
This dependence results from the quadratic term in the model for topological
Weyl semimetals. In weak fields $\omega\ll|\mu|$, $\Delta\Omega_{q}^{\text{b}}$
becomes weakly dependent of the magnetic field. $\Delta\Omega_{q}'$
refers to the special contribution from the chiral Landau band. In
the semiclassical limit, this part is much smaller compared to the
other two parts. Finally, we note that $\Delta\Omega_{q}$ is linear
in $\chi=\hbar^{2}qD^{2}/(8\pi^{2}\rho v_{p}),$ hence the modified
phonon velocity $v_{p}+\Delta v_{p}=v_{p}(1+\Delta\Omega_{q}/v_{p}q)$
remains a constant with respect to $q$. In the weak field limit $\ell_{B}\rightarrow\infty$,
both $\Delta\Omega_{q}^{\text{osc}}$ and $\Delta\Omega_{q}'$ vanish,
whereas $\Delta\Omega_{q}^{\text{b}}$ remains finite and reproduces
the results in the absence of the magnetic field.

Now, we are ready to analyze $\Delta\Omega_{q}$ concretely. In order
to illustrate the results clearly, we consider two complementary cases:
$v_{z}<\sqrt{2}v$ and $v_{z}>\sqrt{2}v$, respectively.

\begin{figure}[htp]
\centering \includegraphics[width=8.3cm]{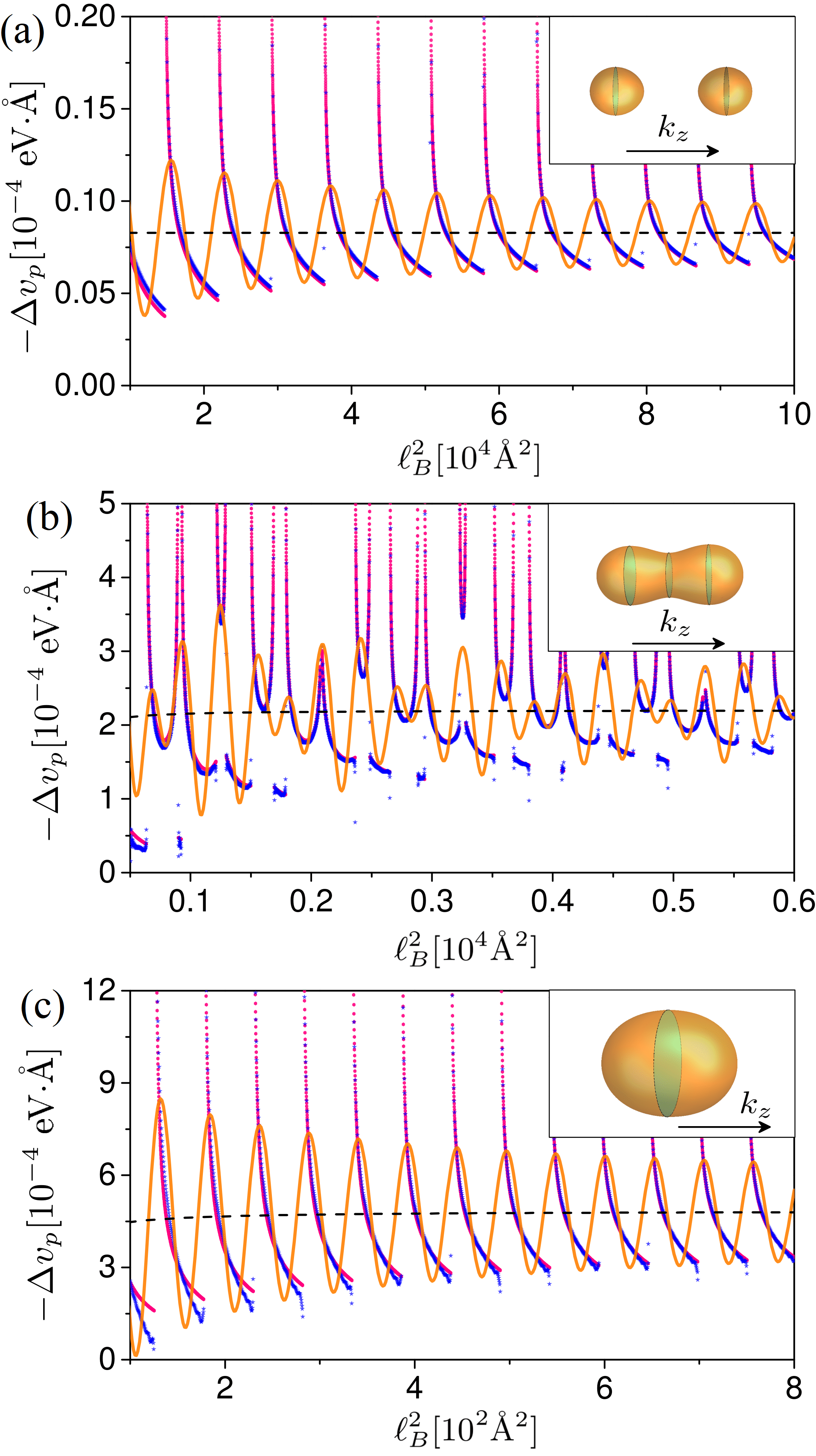} \caption{Modification of the phonon velocity $\Delta v_p$ (or equivalently modification of phonon dispersion $\Delta \Omega_q =\Delta v_pq$) as a function
of $\ell_{B}^{2}$ at $T=0$ and $\mu=0.1$ eV (a), $0.5$ eV (b)
and $1.2$ eV (c), respectively. $\Delta v_p$ is in units of $10^{-4}$ eV$\cdot$$\mathring{\text{A}}=15$ $\hbar\cdot$m/s. The pink dots are calculated numerically
from Eq.\,(\ref{eq:Pi_real}) with $T=0.1$ mK and $\delta=10^{-6}$
eV; the blue dots and yellow solid lines are plots of Eqs.\,(\ref{eq:Pi_real_region1})(a),
(\ref{eq:Pi_real_region2})(b) and (\ref{eq:Pi_real_region3})(c)
with the first 10$^{4}$ harmonics and only the first harmonic, respectively.
The black dashed lines plot the background $\Delta\Omega_{q}^{b}/q$
of oscillations. The insets in each panel are the corresponding Fermi
surfaces with the extremal cross sections in gray color. Other representative parameters
are $v=6$ eV$\cdot$$\mathring{\text{A}}$, $k_{c}=0.15$ $\mathring{\text{A}}$$^{-1}$, $m=15$ eV$\cdot$$\mathring{\text{A}}$$^{2}$,
$\rho=7\times10^{3}$ kg/m$^{3}$, $D=20$ eV, $v_{p}=0.015$ eV$\cdot$$\mathring{\text{A}}$
and $q=10^{-4}$ $\mathring{\text{A}}$$^{-1}$.}

\label{Fig:Pi_real_oscillation}
\end{figure}

\subsection{The case with $v_{z}<\sqrt{2}v$}

If $v_{z}<\sqrt{2}v$, the Weyl semimetal possesses three different
energy regimes, namely, the low- ($|\mu|<E_{\text{L}}$), moderate-
($E_{\text{L}}<|\mu|<vk_{c}$) and high-energy ($|\mu|>vk_{c}$) regimes.
In these regimes, the Fermi surfaces take distinct geometries. While
$E_{\text{L}}$ refers to the transition energy at which the two Fermi
surfaces begin to merge to a single one, $vk_{c}$ is the energy at
which the Fermi surface changes from a dumbbell shape to a convex
sphere. As a result, the phonon dispersion shows different patterns
of oscillations in the magnetic field when $\mu$ is in different
regimes, which we discuss in detail below. For the moment, we consider
$\mu$ to be away from the boundaries between the different energy
regimes. When $\mu$ is close to the boundaries, the variation of
the magnetic field could lead to transitions between different patterns
of oscillations. This is more evident for the opposite case $v_{z}>\sqrt{2}v$
and will discussed later. Figure~\ref{Fig:Pi_real_oscillation}
displays typical $\Delta\Omega_{q}$ as a function of $\ell_{B}^{2}$.
In the plots, we use the typical parameters, e.g., for the Dirac semimetal
Cd$_{3}$As$_{2}$, $v=6$ eV$\cdot$$\mathring{\text{A}}$, eV$\cdot$$\mathring{\text{A}}$, $k_{c}=0.15$
$\mathring{\text{A}}$$^{-1}$, $m=15$ eV$\cdot$$\mathring{\text{A}}$$^{2}$ and $\rho=7\times10^{3}$ kg/m$^{3}$,
and for long wavelength acoustic phonons, $D=20$ eV, $v_{p}=0.015$
eV$\cdot$$\mathring{\text{A}}$ and $q=10^{-4}$ $\mathring{\text{A}}$$^{-1}$\cite{Note2}. The blue dots are the analytic results of Eq.\,(\ref{eq:Pi_real_final})
with the first $10^{4}$ harmonics, while the pink dots are the numerical
integration for Eq.\,(\ref{eq:Pi_real}). They are in perfect agreement
with each other.

We first examine the low-energy limit $\{\eta,\omega\}\ll|\mu|\ll E_{L}$,
where the semimetal possesses two separated pockets of ideal Weyl
fermions located at ${\bf k}=\pm(0,0,k_{c})$, respectively. According
to Eq.\,(\ref{eq:Pi_real_final}), the phonon dispersion modification
$\Delta\Omega_{q}$ can be written as
\begin{eqnarray}
\Delta\Omega_{q} & =- & \dfrac{4\mu^{2}\chi}{v^{2}v_{z}}-\dfrac{4|\mu|\chi}{vv_{z}\ell_{B}}\sum_{j=1}^{\infty}\dfrac{1}{\sqrt{j}}\cos\Big(2j\pi\mathcal{P}_{1}\ell_{B}^{2}-\dfrac{\pi}{4}\Big).\label{eq:Pi_real_ideal}
\end{eqnarray}
The first term is independent of the magnetic field. It is quadratic
in $\mu$ and corresponds to the result in the absence of the magnetic
field. The second term is periodic in $\ell_{B}^{2}$ with the period
$\mathcal{P}_{1}=\mu^{2}/2v^{2}$ quadratic in $\mu/v$ and the amplitude
$4|\mu|/(vv_{\parallel}\ell_{B})$. Unlike the 3D conventional electron
gas, the phase shift (i.e., $-\pi/4$) in the oscillations is always
negative, regardless of the type of carriers. From this phase shift,
a Berry phase of $\pi$ can be deduced, as expected for ideal Weyl
fermions.

Next, we look at the low-energy regime $|\mu|<E_{L}$,
in which the semimetal still has two separated and equivalent
pockets of electrons. In this regime, $\Delta\Omega_{q}$ is formulated
as
\begin{eqnarray}
\Delta\Omega_{q} & = & -\dfrac{\chi|\mu|}{mv}\ln\dfrac{\mathcal{F}_{+}}{\mathcal{F}_{-}}-\dfrac{\chi\mathcal{A}_{1}}{\ell_{B}}\sum_{j=1}^{\infty}\dfrac{1}{\sqrt{j}}\nonumber \\
 &  & \times\cos\Big[2j\pi(\mathcal{P}_{1}\ell_{B}^{2}-\gamma_{1})-\dfrac{\pi}{4}\Big],\label{eq:Pi_real_region1}
\end{eqnarray}
where $\mathcal{P}_{1}=\mu^{2}/2v^{2}$, $\mathcal{F}_{\pm}=v^{2}\pm2m|\mu|+2v\sqrt{m(E_{\text{L}}\pm|\mu|)}$,
$\gamma_{1}=m\mu/v^{2}$, $\mathcal{A}_{1}=4|\mu|/vv_{z}'$ and $v_{z}'=v_{z}\sqrt{1-(\mu/vk_{c})^{2}}$.
The background of the oscillations is written in a logarithmic form.
It depends weakly on the magnetic field, see the black dashed line
in Fig.\,\ref{Fig:Pi_real_oscillation}(a). The period $\mathcal{P}_{1}$
of the oscillations is the same as that of ideal Weyl fermions. This
indicates the relativistic Weyl properties of the two equivalent pockets.
However, compared to ideal Weyl fermions, the Weyl fermions in this
topological semimetal are modified in two aspects. First, the Fermi
velocity $v_{z}'$ is reduced by the Fermi energy as $v_{z}\sqrt{1-(\mu/vk_{c})^{2}}$.
As a result, the amplitude of the harmonic $\mathcal{\chi A}_{1}/\ell_{B}=4|\mu|/vv_{z}'$
is enhanced. Second, the oscillations acquire an additional phase
shift $\gamma_{1}=m\mu/v^{2}$ proportional to $\mu$. Thus, the Berry
phase of the electrons is modified as $\phi_{\text{B}}=\pi(1-2\gamma_{1})=\pi\left(1-2m\mu/v^{2}\right)$.
It is no longer quantized in $\pi$. It decreases for the $n$-type doping
$\mu>0$ whereas increases for the $p$-type doping $\mu<0$.

In experiments, the Berry phase can be measured using the approach
of Landau index analysis. In explicit, we first assigns the ordered indices $\nu=\{1,2,\cdots\}$
to the locations $B_{\nu}$ of the maxima of the oscillations, next
plots $\nu$ as a function of $1/B_{\nu}$ and then extrapolates the
plot to $1/B_{\nu}\rightarrow0$ to get the intercept of $\nu$, say
$\nu_{0}$.  Then, the Berry phase is found via the relation $\phi_{B}=(2\pi\nu_{0}+5\pi/4)~\text{mod }2\pi$.
As mentioned before, the Berry phase for ideal Weyl fermions is $\pi$.
The presence of the quadratic term in the topological semimetal gives
rise to a correction of the Berry phase $2\pi\gamma_{1}=-2\pi m\mu/v^{2}$.
For Cd$_{3}$As$_{2}$ with typical parameters: $k_{c}\simeq0.15$
$\mathring{\text{A}}$$^{-1}$, $v\simeq5\text{\ensuremath{-}}10$ eV$\cdot$$\mathring{\text{A}}$ (i.e., $v/\hbar\simeq0.75\text{\ensuremath{-}}1.5\times10^{6}$
m/s)\ \citep{Liu14natmat,Borisenko14prl,Neupane14nc,Jeon14natmat},
$m\simeq15$ eV$\cdot$$\mathring{\text{A}}$$^{2}$\ \citep{Cano2017PRB}, and in the
low-energy regime, the Fermi energy $\mu$ can be as large as the
Lifshitz transition energy $\simeq0.3$ eV. Thus, the correction can
be as large as $\simeq0.1\text{\ensuremath{-}}0.4\pi$ . For Na$_{3}$Bi
with typical parameters: $k_{c}\simeq0.1$ $\mathring{\text{A}}$$^{-1}$, $v\simeq2.5$
eV$\cdot$$\mathring{\text{A}}$\ \citep{Liu14sci,Xiong15sci}, $m\simeq10$ eV$\cdot$$\mathring{\text{A}}$$^{2}$\ \citep{Wang12prb}
and $\mu\simeq0.1$ eV, we can obtain a correction up to $\simeq0.3\pi$.
This correction may account for the discrepancy between the observations
of the Berry phase in the $n$-type doped Dirac semimetals and the ideal
value of $\pi$\ \citep{He14prl,Narayanan15prl}.

\begin{figure}[h]
\centering

\includegraphics[width=8.5cm]{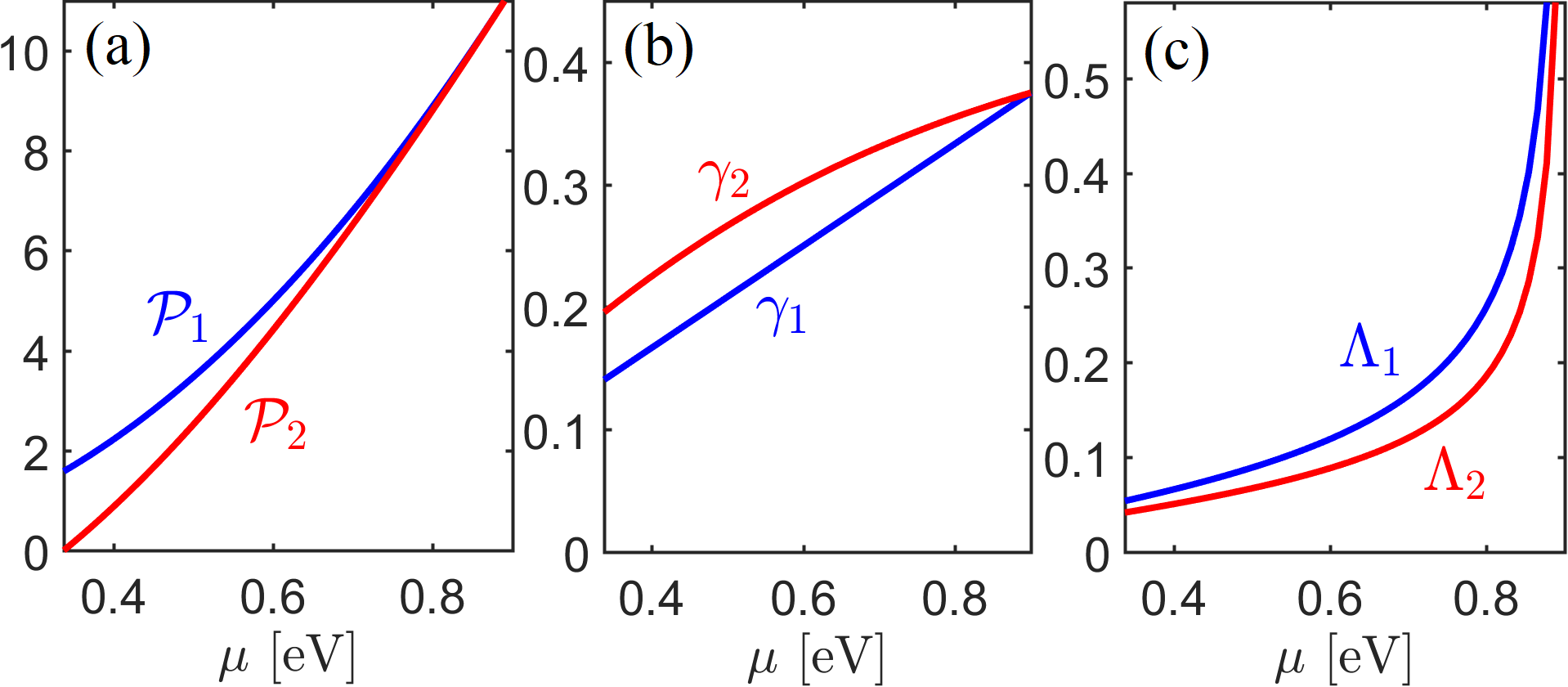} \caption{The periods $\mathcal{P}_{1,2}$ (in units of $10^{3}\mathring{\text{A}}^{2}$)
(a), phase shifts $\gamma_{1,2}$ (in units of $\pi$) (b) and amplitudes
$\Lambda_{1,2}$ (in units of $\text{eV}^{-1}$) (c) of the two harmonics
in $\Delta\Omega_{q}$ as functions of $\mu$ in the moderate-energy
regime. Other parameters are the same as those in Fig.\ \ref{Fig:Pi_real_oscillation}.}

\label{Periods}
\end{figure}

In the moderate-energy regime $E_{L}<|\mu|<vk_{c}$,
there is only a single Fermi surface in the semimetal. It takes a
dumbbell shape with two equivalent maximum cross sections at around
$k_{z}=\pm k_{c}$ as well as a minimum cross section at $k_{z}=0$,
see the inset of Fig.~\ref{Fig:Pi_real_oscillation}(b). Thus, $\Delta\Omega_{q}$
has two harmonics and exhibits a beating pattern of oscillations,
as shown in Fig.~\ref{Fig:Pi_real_oscillation}(b). The first harmonic
is associated with the two maximum cross sections, similar to that
in the low-energy regime, whereas the second one is associated with the minimum
cross section. The corresponding $\Delta\Omega_{q}$ reads
\begin{eqnarray}
\Delta\Omega_{q} & = & -\dfrac{\chi|\mu|}{mv}\ln\dfrac{\mathcal{F}_{+}}{\mathcal{F}_{0}}-\dfrac{\chi}{\ell_{B}}\sum_{\kappa\in\{1,2\}}\sum_{j=1}^{\infty}\dfrac{\mathcal{A}_{\kappa}}{\sqrt{j}}\nonumber \\
 &  & \times\cos\Big[2j\pi\big(\mathcal{P}_{\kappa}\ell_{B}^{2}-\gamma_{\kappa}\big)+(-1)^{\kappa}\dfrac{\pi}{4}\Big],\label{eq:Pi_real_region2}
\end{eqnarray}
where $\mathcal{F}_{0}=\sqrt{v^{4}-4v^{2}mE_{\text{L}}+4m^{2}\mu^{2}}$.
The first harmonic has the same forms of the period $\mathcal{P}_{1}$,
phase shift $\gamma_{1}$ and amplitude $\mathcal{A}_{1}$ as those
in the low-energy regime. In contrast, the period, phase shift and
amplitude of the second harmonic behave more complicatedly. They can
be written as $\mathcal{P}_{2}=(2mE_{\text{L}}-v^{2}+\mathcal{F}_{0})/4m^{2}$,
$\gamma_{2}=m\mu/\mathcal{F}_{0}$ and $\mathcal{A}_{2}=2|\mu|/\sqrt{\mathcal{F}_{0}|v^{2}-\mathcal{F}_{0}|}$,
respectively. Figure~\ref{Periods} displays the periods $\mathcal{P}_{1,2}$,
phase shifts $\gamma_{1,2}$ and amplitudes $\mathcal{A}_{1,2}$ of
the two harmonics as functions of $\mu$. Unlike $\mathcal{P}_{1}\propto\mu^{2}$,
$\mathcal{P}_{2}$ deviates from the quadratic dependence on $\mu$,
which signifies the non-relativistic nature of the associated electrons
centered at $k_{z}=0$. Similar to $\gamma_{1}$, $\gamma_{2}$ increases
as increasing $|\mu|$. However, it is always larger than $\gamma_{1}$.
This indicates a larger deviation of the Berry phase from $\pi$ and
hence provides another signature of the non-relativistic property
of the second harmonic. A similar beating oscillation occurs in the
Shubnikov-de Haas oscillation due to the topological band inversion
even in the absence of Zeeman splitting\ \citep{wang-16xxx}. The
double-period Shubnikov-de Haas oscillation has been experimentally
observed on Cd$_{3}$As$_{2}$~\citep{Zhao15prx}.

Finally, we analyze the high-energy regime $|\mu|>vk_{c}$,
where the Fermi surface becomes a convex spheroid centered at ${\bf k}=0$.
$\Delta\Omega_{q}$ again oscillates with a single period and can
be formulated as
\begin{eqnarray}
\Delta\Omega_{q} & = & -\dfrac{\chi|\mu|}{mv}\ln\dfrac{\mathcal{F}_{+}}{\mathcal{F}_{0}}-\dfrac{\chi\mathcal{A}_{2}}{\ell_{B}}\sum_{j=1}^{\infty}\dfrac{1}{\sqrt{j}}\nonumber \\
 &  & \times\cos\Big[2\pi j\big(\mathcal{P}_{2}\ell_{B}^{2}-\gamma_{2}\big)-\dfrac{\pi}{4}\Big].\label{eq:Pi_real_region3}
\end{eqnarray}
The oscillations are contributed purely by the second harmonic. This
implies that the carriers in the electronic system behave more like
a non-relativistic electron gas. When increasing $|\mu|$, the period $\mathcal{P}_{2}$
increases, while the phase shift $\gamma_{2}$ approaches the constant
$\text{sgn}(\mu)\pi/2$ and the amplitude $\mathcal{A}_{2}$ to $1/m$.
In the extreme limit $|\mu|\gg vk_{c}$, $\mathcal{A}_{2}\approx1/m$,
$\mathcal{F}_{0}=2m|\mu|$, $\mathcal{P}_{2}=|\mu|/2m$ and $\gamma_{2}=\text{sgn}(\mu)/2$.
Equation~(\ref{eq:Pi_real_region3}) resembles the results for a
3D conventional electron gas with parabolic dispersion.

Before closing this subsection, let us discuss the amplitude
of the oscillations. The phonon dispersion modification could be found
by measuring the change of the phonon velocity compared to the unperturbed
one, $\Delta\Omega_{q}/\Omega_{q}=\Delta v_{p}/v_{p}$. The ratio
is determined by $\chi\mathcal{A}_{1,2}/(\Omega{}_{q}\ell_{B})$.
For the typical parameters for the acoustic phonons in the Dirac semimetal
Cd$_{3}$As$_{2}$, $v_{p}=0.015$ eV$\cdot$$\mathring{\text{A}}$, $\rho=7000$ kg/m$^{3}$\ \citep{HaitaoWang07MT},
$D=20$ eV\ \citep{Gerin78PRB} and $q=10^{-4}$ $\mathring{\text{A}}$$^{-1}$, we obtain
$\chi\approx3.3\times10^{-5}$ eV$^{2}\cdot$$\mathring{\text{A}}$$^{3}$. Hence, we estimate
the phonon dispersion modification as $10^{-4}\text{\ensuremath{-}}10^{-3}$
for $\mu\approx0.1$ eV and $\ell_{B}\approx10^{3}$ $\mathring{\text{A}}$. This is experimentally
detectable, for example, by the pulse-echo technique\ \citep{Truell-book,Luthi-book}.
A finite scattering lifetime (or broadening) is usually present in the electronic bands, due to impurity scattering or electron-phonon interaction~\citep{Shoenberg1984}.
In the semiclassical regime, we can assume a constant lifetime $\tau_q$ for the Landau bands. Then, we can derive a Dingle damping factor $e^{-j \pi m^{*}/(eB\tau_{q}) }$ for the $j$-th harmonic, where $m^*$ is the effective mass. The Dingle factors make the first harmonic dominant whereas higher harmonics become less important. However, in a clean Weyl metal, the scattering lifetime $\tau_q$ is very long while the mass $m^*$ is small. Therefore, we could expect to observe a large number of harmonics at low temperatures.

\begin{figure}[h]
\centering

\includegraphics[width=8.5cm]{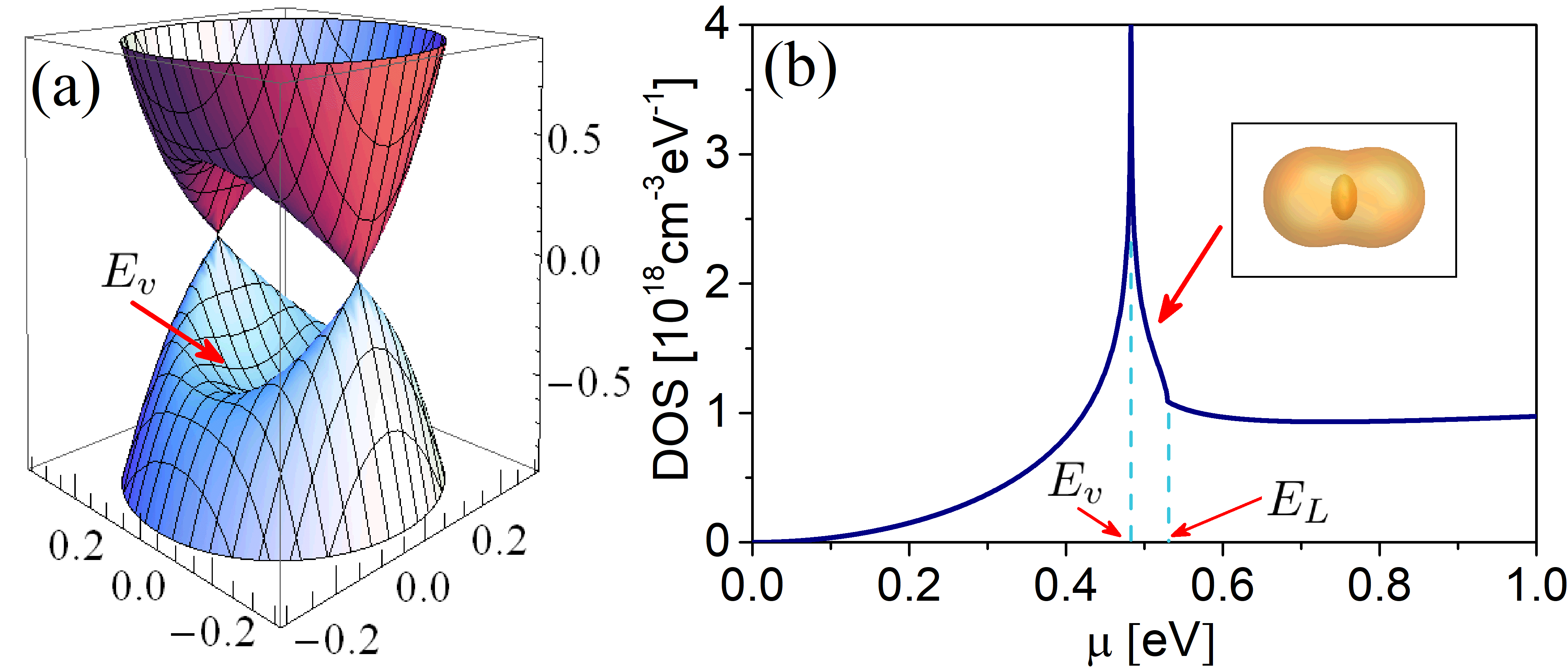}\caption{(a) Band structure and (b) density of states of the topological Weyl
semimetal in the case with $v_{z}>\sqrt{2}v$. A van Hove singularity
happens at $E_{\text{v}}=v\sqrt{v_{z}^{2}-v^{2}}/2m$. The insert
of (b) shows the Fermi surface in the energy regime $E_{\text{v}}<\mu<\text{min}(E_{\text{L}},vk_{c})$.
The parameters are $k_{c}=0.23$ $\mathring{\text{A}}$$^{-1}$, $m=10$ eV$\cdot$$\mathring{\text{A}}$$^{2}$
and $v=2.5$ eV$\cdot$$\mathring{\text{A}}$.}

\label{van_Hove}
\end{figure}

\subsection{The case with $v_{z}>\sqrt{2}v$}

\subsubsection{Oscillations in four energy regimes}

\begin{figure}[h]
\includegraphics[width=8.3cm]{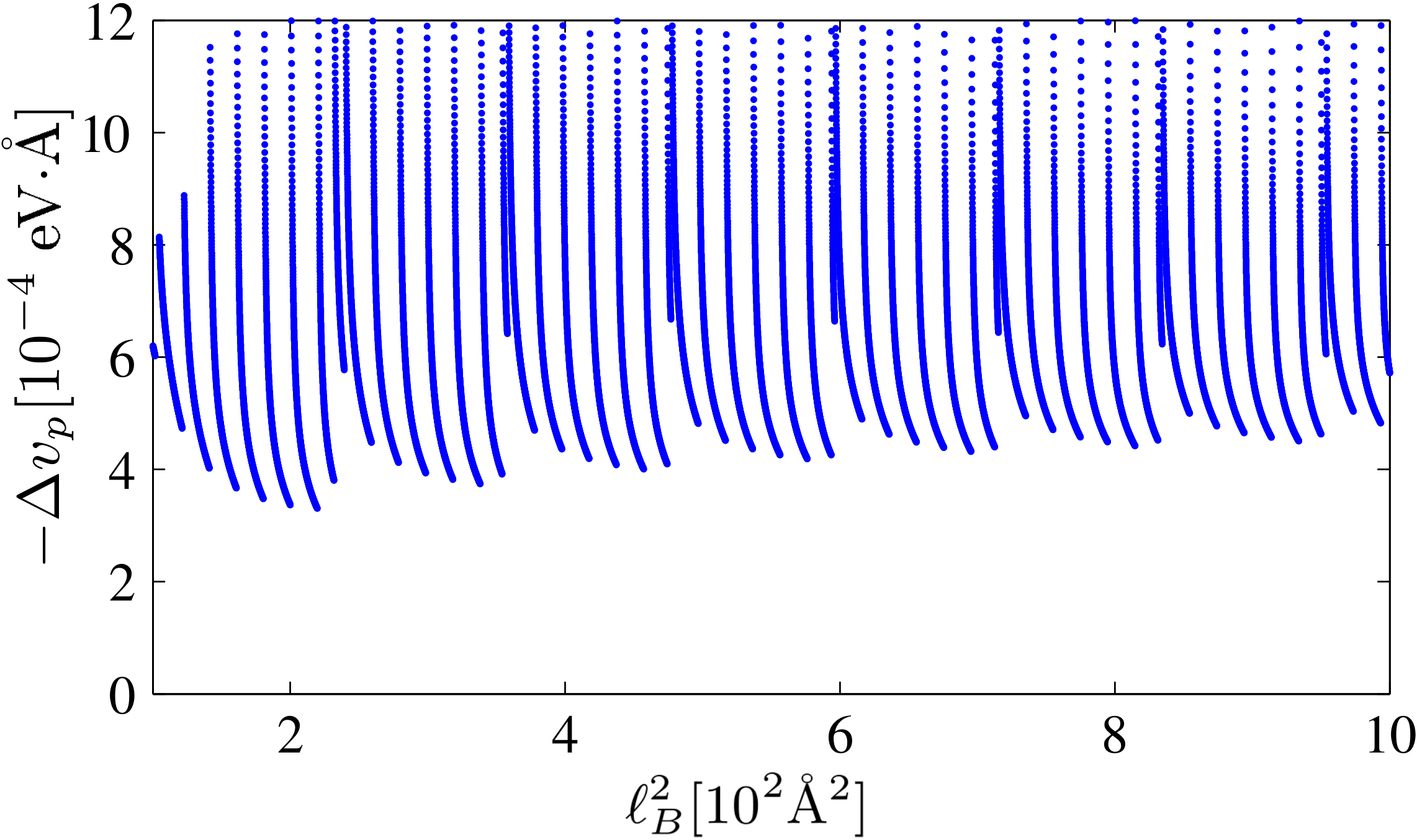}\caption{Modification of the phonon velocity $\Delta v_p$ (or equivalently $\Delta \Omega_q =\Delta v_pq$) as a function of $\ell_{B}^{2}$ for $v<mk_{c}$. $\Delta v_p$ is in units of $10^{-4}$ eV$\cdot$$\mathring{\text{A}}=15$ $\hbar\cdot$m/s.
The beating oscillations have two periods $\mathcal{P}_{2}$ and $\mathcal{P}_{3}$,
different from that in Fig.\ \ref{Fig:Pi_real_oscillation}(b) the
periods of which are $\mathcal{P}_{1}$ and $\mathcal{P}_{2}$. $\mu=0.8$
eV, $k_{c}=0.3$ $\mathring{\text{A}}$$^{-1}$ and other parameters are the same as those
in Fig.\,\ref{Fig:transition}.}

\label{fig:PI_real_beating-1}
\end{figure}

Now we turn to the complementary case with $v_{z}>\sqrt{2}v$. The
band structure of the Weyl semimetal in this case is shown in Fig.\,\ref{van_Hove}(a).
In addition to $E_{\textrm{L}}$ and $vk_{c}$, there is another characteristic
energy, the van Hove singularity at $E_{\text{v}}=v\sqrt{v_{z}^{2}-v^{2}}/2m$,
where the DOS diverges; see Fig.\,\ref{van_Hove}(b). Therefore,
we can define four different energy regimes separated by $E_{\text{v}}$, $E_{\text{L}}$,
and $vk_{c}$. Suppose $\mu$ is away from the boundaries. $\Delta\Omega_{q}$
shows distinct patterns of oscillations in these four regimes, similar
to the case with $v_{z}<\sqrt{2}v$. The low-energy regime $|\mu|<E_{\text{v}}$
is exactly the same as that for the case with $v_{z}<\sqrt{2}v$.
There are two separated and equivalent convex Fermi surfaces at $(0,0,\pm k'_{c})$
where $k'_{c}=\pm\sqrt{k_{c}^{2}-\mu^{2}/v^{2}}$. Thus, $\Delta\Omega_{q}$
oscillates with a single period and can be formulated by Eq.\,(\ref{eq:Pi_real_region1}).
In the second energy regime $E_{\text{v}}<|\mu|<\text{min}(E_{\text{L}},vk_{c})$,
the two separated Fermi surfaces merge to form a dumbbell-shaped Fermi
surface with a sphere inside; see the insert in Fig.\,\ref{van_Hove}(b).
Note that this is different from the case with $v_{z}<\sqrt{2}v$
where the two Fermi surfaces merge at $E_{\text{L}}$ rather than
$E_{\text{v}}$ and no sphere appears inside. Thus, the dumbbell-shaped
Fermi surface has an additional maximal cross section for the inside
sphere. As a result, $\Delta\Omega_{q}$ displays beating oscillations
with three periods, see the small field region in Fig.\,\ref{Fig:transition}(a).
In the third energy regime $\text{min}(E_{\text{L}},vk_{c})<|\mu|<\text{max}(E_{\text{L}},vk_{c})$,
$\Delta\Omega_{q}$ also shows a beating pattern of oscillations but
with only two periods. If $E_{\text{L}}<|\mu|<vk_{c}$, then the inside
Fermi surface disappears and the outside Fermi surface remains dumbbell
shaped. The formula for $\Delta\Omega_{q}$ is the same as Eq.\ (\ref{eq:Pi_real_region2})
with the periods given by $\mathcal{P}_{1}$ and $\mathcal{P}_{2}$.
On the other hand, if $vk_{c}<|\mu|<E_{\text{L}}$, then the outside
Fermi surface becomes convex but it coexists with the inside Fermi
surface. Thus, the two periods of the oscillations are instead given
by $\mathcal{P}_{2}$ and $\mathcal{P}_{3}$, where $\mathcal{P}_{3}=(2mE_{\text{L}}-v^{2}-\mathcal{F}_{0})/4m^{2}$.
The typical oscillations in this case are shown in Fig.\,\ref{fig:PI_real_beating-1}.
Finally, in the high-energy regime $|\mu|>\text{max}(E_{\text{L}},vk_{c})$,
the Fermi surface becomes a single convex sphere. We have again simple
oscillations with only one period and formulated by Eq.\,(\ref{eq:Pi_real_region3}).

\subsubsection{Van Hove singularity and transitions of oscillation patterns}

\begin{figure}[h]
\centering

\includegraphics[width=8.5cm]{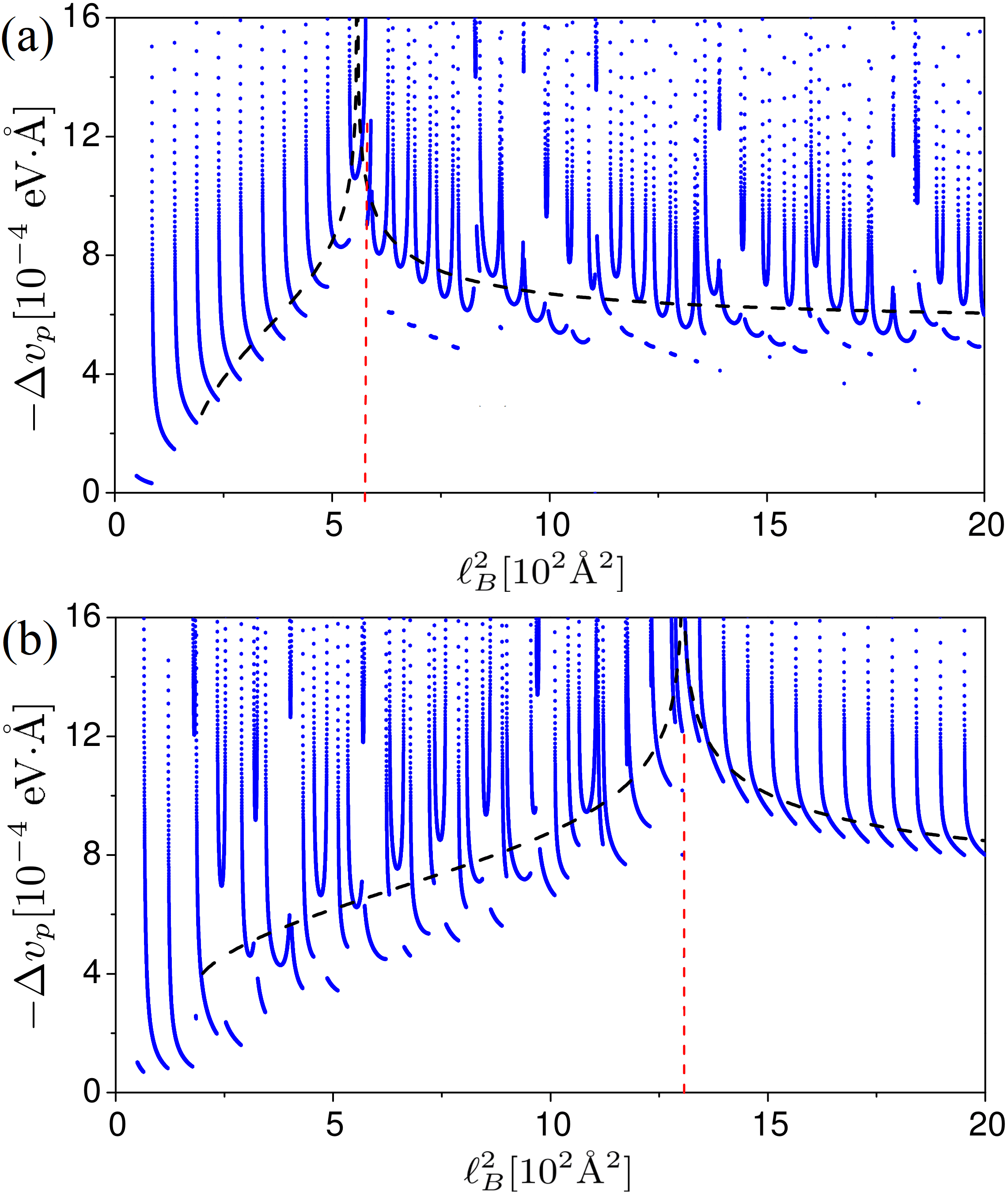}

\caption{Transition between different patterns of oscillations.
$\Delta v_p$ is in units of $10^{-4}$ eV$\cdot$$\mathring{\text{A}}=15$ $\hbar\cdot$m/s.
The blue dots are numerical results from Eq.\,(\ref{eq:Pi_real}).
(a) is for $\mu=0.5$ eV. In the large field (i.e., small $\ell_{B}$) region, $\Omega_{q}$
shows single-periodic oscillations, whereas in the small field (i.e.,
large $\ell_{B}$) region, it shows beating oscillations with three
periods. (b) is for $\mu=-0.475$ eV. In the small field region, $\Omega_{q}$
shows single-periodic oscillations, where in the large field region,
it shows beating oscillations with three periods. The black dashed
lines are the plots of $\Delta\Omega_{q}^{\text{b}}$ stated in Eq.\,(\ref{eq:Pi_real_background}).
The dashed red vertical lines mark the transition of the patterns.
Other parameters are $k_{c}=0.23$
$\mathring{\text{A}}$$^{-1}$, $m=10$ eV$\cdot$$\mathring{\text{A}}$$^{2}$, $v=2.5$ eV$\cdot$$\mathring{\text{A}}$, $D=10$ eV, $v_{p}=0.015$
eV$\cdot$$\mathring{\text{A}}$, $\rho=7000$ kg/m$^{3}$ and $q=10^{-4}$ $\mathring{\text{A}}$$^{-1}$.}

\label{Fig:transition}
\end{figure}

When the Fermi energy $\mu$ is close to the van Hove singularity,
we can observe a transition of different patterns of oscillations
when increasing the magnetic field. To show this, we consider $E_{\text{v}}\apprle\mu<\text{min}(E_{L},vk_{c})$
and display the typical oscillations in Fig\ \ref{Fig:transition}(a).
Before the transition, the oscillations have three periods and can
be described by
\begin{eqnarray}
\Delta\Omega_{q}^{\text{osc}} & = & -\dfrac{\chi}{\ell_{B}}\sum_{\kappa\in\{1,2,3\}}\sum_{j=1}^{\infty}\dfrac{\mathcal{A}_{\kappa}}{\sqrt{j}}\nonumber \\
 &  & \times\cos\Big[2j\pi(\mathcal{P}_{\kappa}\ell_{B}^{2}-\gamma_{\kappa})+(-1)^{\kappa}\dfrac{\pi}{4}\Big],
\end{eqnarray}
where $\gamma_{3}=-\gamma_{2}$ and $\mathcal{A}_{3}=2|\mu|/\sqrt{\mathcal{F}_{0}|\mathcal{F}_{0}+v^{2}|}$.
After the transition, the oscillations have only one period and are
described instead by
\begin{eqnarray}
\Delta\Omega_{q}^{\text{osc}} & = & -\dfrac{\chi\mathcal{A}_{1}}{\ell_{B}}\sum_{j=1}^{\infty}\dfrac{1}{\sqrt{j}}\cos\Big[2j\pi(\mathcal{P}_{1}\ell_{B}^{2}-\gamma_{1})-\dfrac{\pi}{4}\Big].
\end{eqnarray}
When the transition happens, the background of the oscillation also
diverges. According to Eq.\,(\ref{eq:Background}), the background
$\Delta\Omega_{q}^{b}$ of oscillations at $\mu\simeq E_{\text{v}}$
can be written as
\begin{eqnarray}
\Delta\Omega_{q}^{\text{b}} & = & -\dfrac{\chi|\widetilde{\mu}|}{vm}\ln\dfrac{\mathcal{F}'_{+}\mathcal{F}'_{-}}{4m^{2}|\widetilde{\mu}^{2}-E_{\text{v}}^{2}|},\label{eq:Pi_real_background}
\end{eqnarray}
where $\mathcal{F}'_{\pm}=\pm v^{2}+2m|\widetilde{\mu}|+2v\sqrt{m(E_{\text{L}}\pm|\widetilde{\mu}|)}$.
It takes the same form of the DOS but with the Fermi energy modified by
the magnetic field $\widetilde{\mu}=\mu-\omega$. The divergence of
$\Delta\Omega_{q}^{\text{b}}$ happens at $|\widetilde{\mu}|=E_{\text{v}}.$
Thus, the critical magnetic field is given by $B_{c}=(\hbar/me)|E_{\text{v}}-|\mu||$.
The replacement $\widetilde{\mu}=\mu-\omega$ in Eq.\,(\ref{eq:Pi_real_background})
indicates that the transition induced by the magnetic field stems
from the quadratic term in Eq.\,$\left(\ref{eq:model_hamiltonian}\right)$.
A similar transition happens for $\text{min}(E_{L},vk_{c})<\mu\apprle-E_{\text{v}}$
but with the two patterns exchanging their sides, see Fig.\ \ref{Fig:transition}(b).
We note finally that the transition of oscillation patterns is a manifestation
of the van Hove singularity which is absent in the case with $v_{z}<\sqrt{2}v$.

\section{Phonon attenuation\label{sec:Phonon-attenuation-for}}

In this section, we discuss the phonon attenuation. Using Migdal's
theorem, the phonon attenuation $\Gamma_{q}$ is given by Eq.\,(\ref{eq:Pi_imag}).
At low temperatures $T\ll\{\eta,\omega\}$, and for small broadening
of Landau bands, $\Gamma_{q}$ exhibits an oscillatory and striking
spike-like pattern, called giant quantum oscillations~\citep{Gurevich61JETP}, as
the magnetic field varies. These peaks stem from the resonant absorption
of the acoustic phonons. The spikes occur periodically when the Fermi
energy $\mu$ falls near the edge of one Landau band and the small
component of the Fermi velocity in the field direction equals the
phonon velocity.

To facilitate the analysis of the phonon attenuation, we first analyze
the zero-temperature limit and then consider the effect of a low temperature.
The phonon attenuation at zero temperature is given by Eq.\,(\ref{eq:Pi_attentuation}).
In the semiclassical limit, we apply Poisson's summation rule to
Eq.\,(\ref{eq:Pi_attentuation}) and rewrite it as
\begin{eqnarray}
\Gamma_{q}^{0}= & \dfrac{\pi\chi v_{p}}{\ell_{B}^{2}} & \sum_{s}\int_{0}^{\infty}dn\int dk_{z}\delta(\epsilon_{k_{z}}^{ns}-\mu)\nonumber \\
 & \times & \delta\big(v_{k_{z}}^{ns}+v_{p}\big)\Big[1+2\sum_{j=1}^{\infty}\cos(2\pi jn_{k_{z}})\Big].
\end{eqnarray}
Similar to the previous section for the modification of phonon dispersion
$\Delta\Omega_{q}$, we have focused on the case with $\theta=0$.
Integrating over $n$ and $k_{z}$, this yields
\begin{eqnarray}
\Gamma_{q}^{0} & = & \pi\chi v_{p}|\widetilde{\mu}|\sum_{s,l}\sum_{k_{i}^{l}}\Big\{\dfrac{\Theta(n_{l,k_{z}})}{\mathcal{F}_{k_{z}}|\partial^{2}\epsilon_{k_{z}}^{ns}/\partial k_{z}^{2}|_{n_{l}}}\nonumber \\
 & ~ & ~\ \ \ \ \ \ \ \ \ \times[1+2\sum_{j=1}^{\infty}\cos(2\pi jn_{l,k_{z}})]\Big\}_{k_{i}^{l}},\label{eq:GammaT0}
\end{eqnarray}
where $n_{l,k_{z}}$ with $l=\pm$ is given by Eq.\,(\ref{eq:n-function});
the sum runs over all possible solutions $k_{i}^{l}$ to $v_{k_{z}}^{n_{l}s}=0$.
The phonon attenuation $\Gamma_{q}^{0}$ is linear in the momentum
of phonons $q$, but independent of the velocity $v_{p}$. The oscillations
$\Gamma_{q}^{0}$ are also determined essentially by the extremal
cross sections of the Fermi surface and have the same periods in the
oscillations of $\Delta\Omega_{q}$. Therefore, the measurement of
phonon attenuation provides an alternative way to detect the morphology
of the Fermi surface. In contrast to $\Delta\Omega_{q}$, there is no
suppressing factor of $1/\sqrt{j}$ accompanying each term. Thus,
the higher-order harmonics also make important contributions and give
rise to a series of giant peaks at low temperatures. The positions
of the peaks in the phonon attenuation coincide with that of the singularities
in $\Delta\Omega_{q}$. However, $\Gamma_{q}^{0}$ takes the opposite
sign of $\Delta\Omega_{q}$, which indicates that they are in opposite
phases. It is also noteworthy that, unlike $\Delta\Omega_{q}$, two
$\delta$ functions cancel out in the integrals over $n$ and $k_{z}$.
As a result, there is no phase shift of $\pm\pi/4$ arising from
the saddle point approximation. This may make it clearer to extrapolate
the Berry phase by the index plotting.

\begin{figure}[htp!]
\centering \includegraphics[width=8cm]{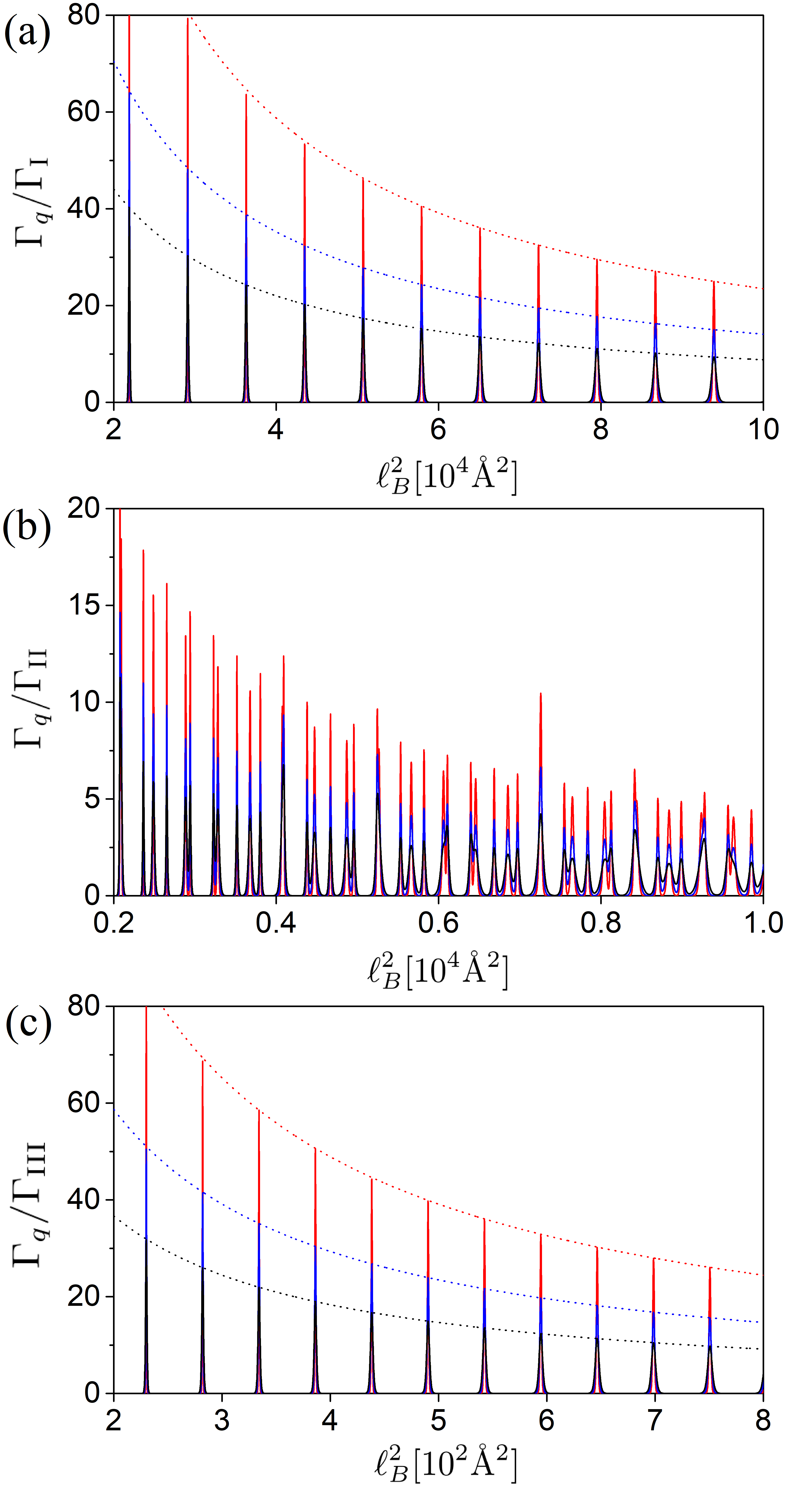}
\caption{Phonon attenuation as a function of $\ell_{B}^{2}$ for $\mu=0.1$
eV (a), $0.5$ eV (b), and $1.2$ eV (c), respectively. The results
are calculated numerically from Eq.\ (\ref{eq:Pi_imag}). The red,
blue and back lines correspond to the temperatures of $3\mu/E_{L}$,
$5\mu/E_{L}$ and $8\mu/E_{L}$ (in units of K), respectively. In
(a) and (c), the dots are the plots of the peak lines in Eqs.\,(\ref{eq:peak-line1})
and (\ref{eq:peak-line2}), respectively. Other parameters are the
same as those in Fig.\,\ref{Fig:Pi_real_oscillation}.}

\label{Fig:giantoscillation}
\end{figure}

The effect of a low temperature on $\Gamma_{q}$ can be taken into
account via the convolution\ \citep{Shoenberg1984}
\begin{eqnarray}
\Gamma_{q} & = & -\int_{-\infty}^{\infty}d\xi\dfrac{\partial f(\xi)}{\partial\xi}\Gamma_{q}^{0}(\xi),\label{Tdependent}
\end{eqnarray}
where the derivative of the Fermi function $\partial f(\xi)/\partial\xi=-\cosh^{-2}[\xi-\mu)/2T]/4T$
is strongly peaked at $\xi=\mu$ with a width determined by the temperature.
Consider a general integral of the form
\begin{eqnarray}
\tilde{\Gamma} & = & \dfrac{1}{4T}\int_{-\infty}^{\infty}d\xi\cosh^{-2}\Big(\frac{\xi-\mu}{2T}\Big)K(\xi)\cos[\phi(\xi)]\label{Tdependent2}
\end{eqnarray}
where $K(\xi)\equiv\Theta(n_{l,k_{z}})/(\mathcal{F}_{k_{z}}|\partial^{2}\epsilon_{k_{z}}^{ns}/\partial k_{z}^{2}|_{n_{l}})$
is a smooth function of $\xi$ and $\phi(\xi)\equiv2\pi jn_{l,k_{z}}$
is also smooth but oscillatory in $\xi$. At low temperatures, the
most important part of the integral arises from the neighborhood of
$\xi=\mu$. Due to the large value $\ell_{B}^{2}$ in $n_{l,k_{z}},$
the term $\cos[\phi(\xi)]$ oscillates rapidly with varying $\xi$.
We expand $\phi(\xi)$ to the first order in $(\xi-\mu)$ and replace
the function $K(\xi)$ by its value at $\xi=\mu$ in Eq.\ (\ref{Tdependent2}).
Then, the integral can be derived to
\begin{eqnarray}
\tilde{\Gamma} & = & -\dfrac{\lambda}{\sinh(\lambda)}K(\mu)\cos[\phi(\mu)],
\end{eqnarray}
where $\lambda=\pi T[\partial\phi(\xi)/\partial\xi]_{\xi=\mu}$. Applying
this approach to all terms with a cosine function in Eq.~(\ref{Tdependent}),
we obtain
\begin{eqnarray}
\Gamma_{q} & = & \pi\chi v_{p}|\widetilde{\mu}|\sum_{s,l}\sum_{k_{i}^{l}}\dfrac{\Theta(n_{l,k_{i}^{l}})}{\mathcal{F}_{k_{z}}|\partial^{2}\epsilon_{k_{z}}^{ns}/\partial k_{z}^{2}|_{n_{l,k_{i}^{l}}}}\nonumber \\
 &  & ~\times\Big[1+2\sum_{j=1}^{\infty}\dfrac{\lambda_{j}^{i,l}\cos(2\pi jn_{l,k_{i}^{l}})}{\sinh(\lambda_{j}^{i,l})}\Big],
\end{eqnarray}
where $\lambda_{j}^{i,l}=2\pi^{2}jT[\partial n_{l,k_{i}^{l}}/\partial\xi]_{\xi=\mu}$.

Now we consider the case with $v_{z}<\sqrt{2}v$ and analyze the features
of $\Gamma_{q}$ in the three different energy regimes. Similar analysis
and main results apply to the other case with $v_{z}>\sqrt{2}v$.
We first look at the low-energy regime $\mu<E_{\text{L}}$. The phonon
attenuation $\Gamma_{q}$ reads
\begin{eqnarray}
\Gamma_{q} & = & \Gamma_{\text{I}}\Big\{1+2\sum_{j=1}^{\infty}\dfrac{\lambda_{j}\cos[2\pi j\big(\mathcal{P}_{1}\ell_{B}^{2}-\gamma_{1}\big)]}{\sinh(\lambda_{j})}\Big\},\label{GiantORegion1}
\end{eqnarray}
where $\lambda_{j}=2\pi^{2}j\ell_{B}^{2}T|\mu|/v^{2}$; $\Gamma_{\text{I}}=\chi v_{p}\mathcal{A}_{1}^{2}/8$
is the magnitude of the phonon attenuation in the absence of the magnetic
field. $\Gamma_{q}$ has the same period $\mathcal{P}_{1}$ as that
stated for $\Delta\Omega_{q}$. In Fig.\,\ref{Fig:giantoscillation}(a),
we plot typical $\Gamma_{q}$ as functions of $\ell_{B}^{2}$ at a
small $\mu(<E_{\text{L}})$ and at three different low temperatures,
respectively. The spike peaks appear periodically and their magnitudes
decay when decreasing the magnetic field or increasing the temperature.
According to Eq.~(\ref{GiantORegion1}), the spike peaks are given
by
\begin{eqnarray}
\Gamma_{q}^{P1} & = & \Gamma_{\text{I}}\Big[1+2\sum_{j=1}^{\infty}\dfrac{\lambda_{j}}{\sinh(\lambda_{j})}\Big].
\end{eqnarray}
At a low temperature $T\ll v^{2}/2\pi^{2}\ell_{B}^{2}|\mu|$, we can
replace the sum over $j$ by an integral and obtain
\begin{eqnarray}
\Gamma_{q}^{P1} & \approx & \dfrac{v^{2}}{4\ell_{B}^{2}T|\mu|}\Gamma_{\text{I}}.\label{eq:peak-line1}
\end{eqnarray}
The magnitude of a peak is proportional to the magnetic field at which
the peak takes place, see the dashed line in Fig.\,\ref{Fig:giantoscillation}(a).
Since $\Gamma_{\text{I}}$ is essentially independent of $T$, the
peak is inversely proportional to the temperature, consistent with
the numerical results presented in Fig.~\ref{Fig:giantoscillation}.

\begin{figure}[htp!]
\centering \includegraphics[width=8.3cm]{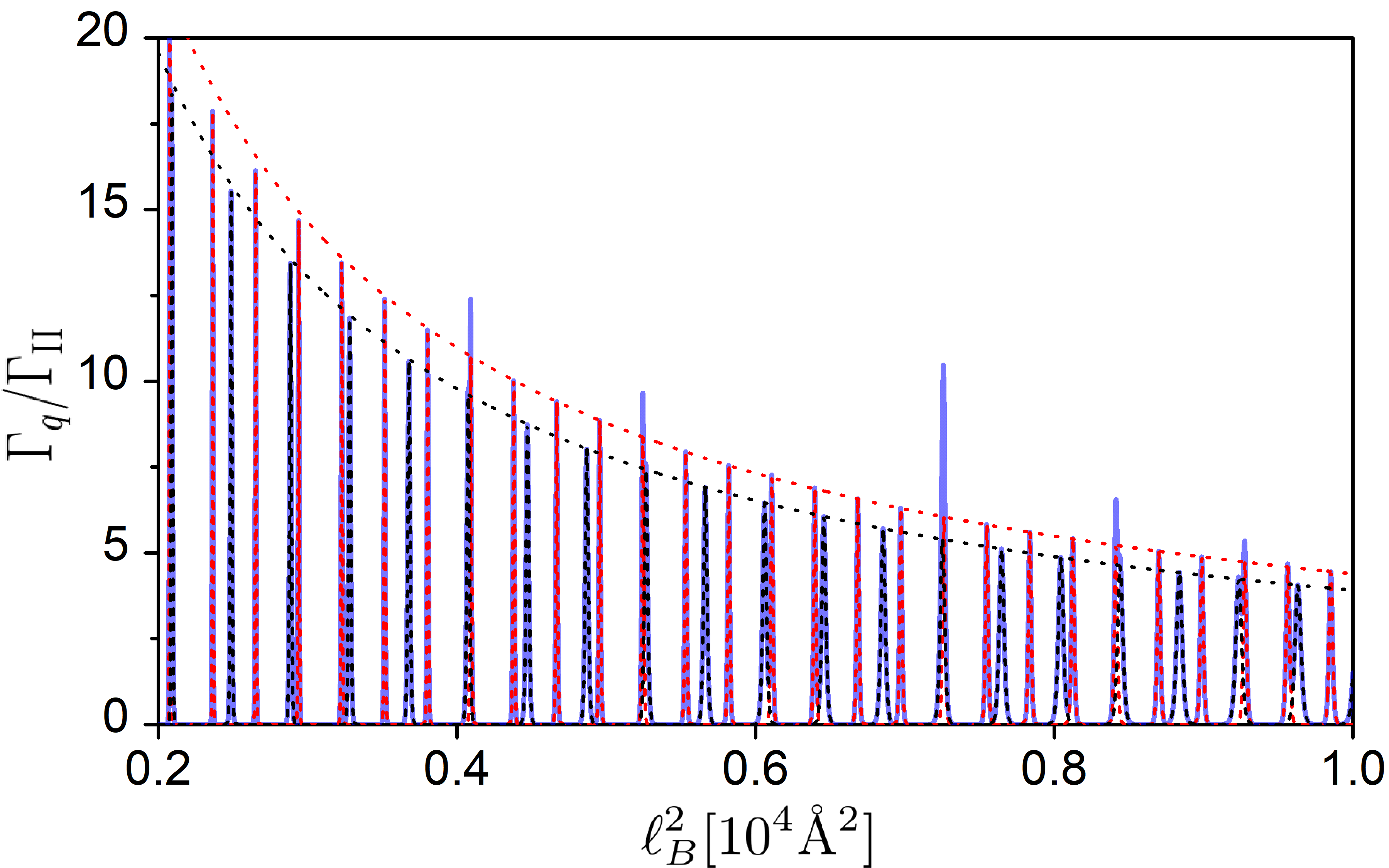}
\caption{The solid blue lines are the same as Fig.\,\ref{Fig:Pi_real_oscillation}(b)
but only for $T=3\mu/E_{L}$. The phonon attenuation can be decomposed
as two different sets of harmonics, as distinguished by the dashed
red and blue lines. The two dotted lines are the plots of the peak
lines in Eqs.\,(\ref{eq:peak-line1}) and (\ref{eq:peak-line2}),
respectively. Other parameters are the same as those in Fig.\ \ref{Fig:Pi_real_oscillation}.}

\label{Fig:giantoscillation2}
\end{figure}

To study the shape of the peaks, we turn to Eq.\,(\ref{eq:Pi_imag})
and write approximately the phonon attenuation as
\begin{eqnarray}
\Gamma_{q} & = & \frac{\pi\chi v_{p}}{4l_{B}^{2}T}\sum_{n}\frac{\sqrt{n}\eta}{k_{c}^{2}-n/\ell_{B}^{2}}\cosh^{-2}\left(\frac{\sqrt{n}\eta-\ell_{n}^{2}}{2T}\right).\label{eq:peaks}
\end{eqnarray}
A specific peak is described by a corresponding term in the summation
of Eq.~(\ref{eq:peaks}). In general, the shape of a peak can be
described by
\begin{eqnarray}
\Gamma_{n} & = & A_{n}\cosh^{-2}[|\mu|(\ell_{B}^{2}-\ell_{n}^{2})/(2\ell_{n}^{2}T)],\label{eq:peak-shape}
\end{eqnarray}
where $\ell_{n}^{2}$ corresponds to the magnetic field where the
center of the peak locates; the magnitude $A_{n}$ can be taken as
a constant with respect to $\ell_{B}$ around the peak. From Eq.\,(\ref{eq:peak-shape}),
the half width of the peak is found as
\begin{eqnarray}
(\delta\ell_{B}^{2})_{n} & = & 3.53\ell_{n}^{2}T/|\mu|=1.76\ell_{n}^{2}T\gamma_{2}/(m\mathcal{P}_{2}).\label{eq:m-relation}
\end{eqnarray}
It is linear in $T$ and $\ell_{B}^{2}$, as shown in Fig.\,\ref{Fig:giantoscillation}.

In the moderate-energy regime, the phonon attenuation consists
of two different periods:
\begin{eqnarray}
\Gamma_{q} & = & \Gamma_{\text{II}}\Big\{\zeta_{1}\Big[1+2\sum_{j=1}^{\infty}\dfrac{\lambda_{j}\cos[2\pi j\big(\mathcal{P}_{1}\ell_{B}^{2}-\gamma_{\text{1}}\big)]}{\sinh(\lambda_{j})}\Big]\nonumber \\
 &  & +\zeta_{2}\Big[1+2\sum_{j=1}^{\infty}\dfrac{\lambda_{j}'\cos[2\pi j\big(\mathcal{P}_{2}\ell_{B}^{2}-\gamma_{2}\big)]}{\sinh(\lambda_{j}')}\Big]\Big\},\label{GiantORegion2}
\end{eqnarray}
where $\lambda_{j}'=2\pi^{2}j\ell_{B}^{2}T|\mu|/\mathcal{F}_{0},$
$\Gamma_{\text{II}}=\chi v_{p}(\mathcal{A}_{1}^{2}+2\mathcal{A}_{2}^{2})/8$
with $\zeta_{1}=\mathcal{A}_{1}^{2}/(\mathcal{A}_{1}^{2}+2\mathcal{A}_{2}^{2})$
and $\zeta_{2}=2\mathcal{A}_{2}^{2}/(\mathcal{A}_{1}^{2}+2\mathcal{A}_{2}^{2})$
is the phonon attenuation in the absence of magnetic field. These
two periods, analogous to the previous argument for $\Delta\Omega_{q}$,
can be attributed to the Weyl and non-relativistic electrons, respectively.
Equipped with Eq.~(\ref{GiantORegion2}), we can identify the complicated
pattern of the phonon attenuation in Fig.~\ref{Fig:giantoscillation}(b)
as a combination of two simple sets of spike peaks with different
periods and magnitudes, as shown in Fig.~\ref{Fig:giantoscillation2}.
One set of peaks associating with the Weyl electrons is still described
by Eq.~(\ref{eq:peak-line1}). The other set of peaks is related
to non-relativistic electrons and can be instead depicted by
\begin{eqnarray}
\Gamma_{q}^{P2} & = & \dfrac{\chi v_{p}\mathcal{A}_{2}^{2}}{4|\mu|}\dfrac{\mathcal{F}_{0}}{4\ell_{B}^{2}T}.\label{eq:peak-line2}
\end{eqnarray}
Similar to $\Gamma_{q}^{P1}$, $\Gamma_{q}^{P2}$ are linear in the
magnetic field and inversely proportional to the temperature. The
half width for these peaks can also be written in the form
\begin{eqnarray}
(\delta\ell_{B}^{2})_{n} & = & 1.76\ell_{n}^{2}T\gamma_{2}/(m\mathcal{P}_{2}).\label{eq:m-relation2}
\end{eqnarray}
Note that in this energy regime, the peaks are not always well separated by broad minima. Two peaks from different
sets may merge to be a single enhanced peak; see the peak around $\ell_{B}^{2}\sim0.7\times10^{4}$$\mathring{\text{A}}$$^{2}$
in Fig.~\ref{Fig:giantoscillation2}.

Finally, we look at the high-energy regime. The phonon attenuation
oscillates with only the second period and reads
\begin{eqnarray}
\Gamma_{q} & = & \Gamma_{\text{III}}\Big[1+2\sum_{j=1}^{\infty}\dfrac{\lambda_{j}'\cos\big(2\pi j\big(\mathcal{P}_{2}\ell_{B}^{2}-\gamma_{2}\big)\big)}{\sinh(\lambda_{j}')}\Big].
\end{eqnarray}
where $\Gamma_{\text{III}}=\chi v_{p}\mathcal{A}_{2}^{2}/4$. It associates
with the non-relativistic electron pocket and the corresponding set
of peaks is given by Eq.~(\ref{eq:peak-line2}). In the extreme limit
$|\mu|\gg vk_{c}$, the peaks become
\begin{equation}
\Gamma_{q}^{P2}(\mu,T) = \chi v_{p}/(8m\ell_{B}^{2}T),\label{eq:peak-line3}
\end{equation}
which is independent of the Fermi energy $\mu$.

Let us end this section by discussing some experimental relevance
for the phonon attenuation. For the Dirac semimetal Cd$_{3}$As$_{2}$
with typical parameters $v\simeq5\text{\ensuremath{-}10}$ eV$\cdot$$\mathring{\text{A}}$,
and for $\mu\simeq0.1$ eV and $B\simeq1$ T, a comparable ratio between
the half width and the period $(\delta\ell_{B}^{2})_{n}\sim1/\mathcal{P}_{1}$
yields the temperature $T\sim0.57v^{2}/\ell_{B}^{2}|\mu|\sim2\text{\ensuremath{-}}9\times10^{-3}$
eV (i.e., 25-100 K). Thus, the spike peaks could be observed at temperatures
on the order of 1K. The electron scattering would broaden the electron
bands, and obscure the $k_{z}$ of absorbing electrons at the band
edges. This would further broaden the spike peaks. Thus, to observe
the giant oscillations, the broadening of peaks needs to be much
smaller than the separation between the peaks. This requirement turns out as $ql \gg 1$, where $l = v\tau_q$ is the mean
free length and $\tau_q$ the quantum scattering lifetime
of electrons. These conditions could be simultaneously
satisfied in the Dirac semimetals Cd$_{3}$As$_{2}$, Na$_{3}$Bi and the Weyl semimetals TaAs
in which the Shubinkov-de Haas oscillations have been observed\,\citep{He14prl,Zhao15prx,Xiang15PRL,Narayanan15prl,LiH16natcom,Xiong15sci,Xiong16EPL,HuangXC15prx,Luo15PRB,ZhangCL16natcom,WangZ15arXiv}.
Similar to the modification of phonon dispersion, the phonon attenuation
can also be measured in the pulse-echo set-up including the phase-sensitive-detection
technique\ \citep{Truell-book}. Moreover, the spike-like character
of the phonon attenuation at low temperatures allows us to determine
more accurately the field location of the peaks for the index plot.
This constitutes an advantage of the quantum oscillations of phonon
attenuation over other de Haas-van Alphen oscillations.

\section{Summary \label{sec:Summary}}
In summary, we have showed that the energy spectrum of long-wavelength acoustic phonons propagating in a Weyl semimetal exhibits exotic singularity and oscillatory behaviors as the magnetic field varies at low temperatures. These behaviors are very sensitive to the Fermi energy of the Weyl semimetal. When the Fermi energy lies above the Lifshitz transition energy, multiple periods appear in the oscillations, leading to beating behaviors. The van Hove singularity of the Weyl spectrum could manifest as a transition between different patterns of oscillations. We have revealed that the phonon attenuation displays similar oscillatory but spike-peaked behaviors. We have also briefly discussed the experimental relevance of these phonon behaviors in the candidate systems Cd$_{2}$As$_{3}$ and Na$_{3}$Bi.


We thank Shun-Qing Shen and Hai-Zhou Lu for useful discussions. S.B.Z. was supported by the DFG (SPP1666, SFB1170 ``ToCoTronics''),
the W\"urzburg-Dresden Cluster of Excellence ct.qmat, EXC2147, project-id
39085490, and the Elitenetzwerk Bayern Graduate School on ``Topological insulators''.
J.Z. was supported by the 100 Talents Program of Chinese Academy of Sciences (CAS) and by the High Magnetic Field Laboratory of Anhui Province.


\begin{thebibliography}{67}
\bibitem [{footNote2()}]{footNote2}%
  \BibitemOpen
  \bibinfo {note} {When the Fermi energy is close to the Weyl nodes, the Fermi surfaces are almost isotropic. Thus, the results in this low-energy regime should be independent of the direction of the magnetic field. However, when the Fermi energy is far away from the Weyl nodes, then the anisotropy of the electronic structure becomes pronounced. In this case, our theory is not applicable when the direction of magnetic field is deviated far from the $z$ direction.}\BibitemShut {Stop}%
\end{thebibliography}
%

\end{document}